\documentclass[12pt]{article}

\usepackage{scicite}

\usepackage{times}

\newenvironment{sciabstract}{%
\begin{quote} \bf}
{\end{quote}}

\usepackage{amsmath, amssymb}
\usepackage{mathtools}
\usepackage{physics}
\usepackage{placeins}
\usepackage{hyperref}

\title{Temperature-dependent dynamic nuclear polarization of diamond } 

\author
{Gevin von Witte$^{1,2}$, Aaron Himmler$^{2}$, Konstantin Tamarov$^{3}$, \\Jani O. Moilanen$^{4}$, Matthias Ernst$^{2}$,  Sebastian Kozerke$^{1\ast}$\\
\\
\normalsize{$^{1}$Institute for Biomedical Engineering, University and ETH Zurich, 8092 Zurich, Switzerland}\\
\normalsize{$^{2}$Institute of Molecular Physical Science, ETH Zurich, 8093 Zurich, Switzerland}\\
\normalsize{$^{3}$Department of Technical Physics, University of Eastern Finland, 70210 Kuopio, Finland}\\
\normalsize{$^{4}$Department of Chemistry, Nanoscience Center, University of Jyväskylä, 40014 Jyväskylä, Finland}\\
\\
\normalsize{$^\ast$To whom correspondence should be addressed; E-mail: kozerke@biomed.ee.ethz.ch}
}

\date{}

\begin{document} 


\baselineskip24pt


\maketitle

\noindent \textbf{Teaser:} Dynamic nuclear polarization and the involved electrons in diamond change with temperature and microwave power.

\begin{sciabstract}
Dynamic nuclear polarization (DNP) can increase nuclear magnetic resonance (NMR) signals by orders of magnitude. 
DNP in diamond proceeds through different DNP mechanisms with a possible temperature-dependence.
We report on \textsuperscript{13}C dynamic nuclear polarization (DNP) experiments in diamonds at 3.4\,T and 7\,T between 300\,K and 1.7\,K.
Nuclear polarization enhancements between 100 and 600 were measured for all temperatures, corresponding to polarizations at 7\,T between 0.1\% (300\,K) and 38\% (1.7\,K).
A strong temperature dependence of the DNP profiles was observed. 
Longitudinal-detected (LOD) electron paramagnetic resonance (EPR) experiments revealed an additional broad temperature-dependent electron line centered around the $m_I = 0$ line of the P1 triplet transitions.
Our results suggest that nuclei are preferentially polarized via a direct hyperfine mediated polarization transfer while spin diffusion in the sample plays a minor role.

\end{sciabstract}

\section*{Introduction} 
Diamond is one of the most fascinating materials due to its mechanical, optical and thermal properties.
In the last decades, diamond emerged as one of the prime materials for quantum applications due to its long coherence and relaxation times as well as wide variety of defects \cite{wolfowicz_quantum_2021,bradac_quantum_2019}.
The most studied defect in diamond for quantum applications, e.g., information processing, sensing and communication, is the nitrogen-vacancy (NV) center owing to its long coherence time and robust optical initialization and read-out \cite{doherty_nitrogen-vacancy_2013}.
This makes it possible to use electron spins of NV centers as indirectly optically-addressable long-lived spin qubits \cite{bradley_ten-qubit_2019,bartling_entanglement_2022}.
NV centers consist of a substitutional nitrogen atom and a vacancy in the carbon structure. 
Its coherence time is ultimately limited by the electron spin-lattice relaxation time ($T_2\leq2T_1$) \cite{wolfowicz_quantum_2021}, which is affected by neighboring nuclear and electron spin fluctuations \cite{wolfowicz_quantum_2021}.
Decoupling schemes \cite{ryan_robust_2010,de_lange_universal_2010,naydenov_dynamical_2011,zhao_decoherence_2012,cai_robust_2012,xu_coherence-protected_2012,wang_comparison_2012,pham_enhanced_2012,de_lange_controlling_2012,bar-gill_suppression_2012,farfurnik_optimizing_2015}, reduced temperature with decoupling \cite{bar-gill_solid-state_2013}, near unity electron polarization \cite{takahashi_quenching_2008} or isotope engineering \cite{mizuochi_coherence_2009,balasubramanian_ultralong_2009,bradley_ten-qubit_2019,bartling_entanglement_2022} can be employed to extend the coherence time.

NV centers are commonly created from P1 centers.
P1 centers are single substitutional nitrogen atoms within a diamond structure.
P1 centers are spin-1/2 electronic defects and their electron paramagnetic resonance (EPR) spectrum is characterized by hyperfine coupling to the 99.6\% natural abundance \textsuperscript{14}N $I = 1$ nuclear spins, which causes a splitting into three electron lines.
A NV center (electron spin-1 defect) is formed if a P1 center has a diamond lattice vacancy (missing carbon atom) in one of its neighboring lattice sites.
Vacancies can be created during crystal growth or electron irradiation with a P1-to-NV conversion ratio  around 10\%, although values up to 25\% have been reported \cite{mindarava_efficient_2020}.
Thus, each NV center is surrounded by a number of P1 centers, making these usually the most abundant paramagnetic defect and, therefore, a major relaxation source. 
The benefits of high P1 thermal electron polarization have been previously shown to extend the NV coherence time through the suppression of flip-flop spin processes \cite{takahashi_quenching_2008}.

Hyperpolarization of \textsuperscript{13}C nuclear spins in diamond further improves the relaxation time of NV centers \cite{belthangady_dressed-state_2013}.
At room temperature, nuclear polarization levels of 5-6\% have been measured for diamond single crystals when the magnetic field was aligned with the NV axis \cite{king_room-temperature_2015,kavtanyuk_achieving_2024}.
In powder samples with a random orientation of the NV centers relative to the static magnetic field, polarization levels of 0.25\% for 200-250\,\textmu m  \cite{ajoy_orientation-independent_2018} and 0.04\% for 2\,\textmu m \cite{blinder_13c_2024} diamond particles have been achieved.
The prolonged coherence time of hyperpolarized \textsuperscript{13}C nuclei in diamond \cite{sharma_enhancement_2019,beatrez_electron_2023} suggests further applications to improve hyperpolarized nuclear magnetometry \cite{sahin_high_2022} or nuclear spin qubits \cite{bradley_ten-qubit_2019,bartling_entanglement_2022}.

Hyperpolarization of diamond with P1 centers at liquid-helium temperature has been investigated for hyperpolarized nanoparticle magnetic resonance imaging (MRI) applications \cite{rej_hyperpolarized_2015,waddington_nanodiamond-enhanced_2017,kwiatkowski_direct_2018,waddington_phase-encoded_2019}.
Long-lasting nuclear polarizations of a few tens of percent have been achieved at a few Tesla magnetic fields.
The P1 DNP profiles around 3.5-4\,K revealed two broad DNP lobes for either positive or negative enhancement \cite{rej_hyperpolarized_2015,kwiatkowski_direct_2018}. 


Room temperature hyperpolarization of diamond with P1 centers under static \cite{rej_hyperpolarized_2015,shimon_large_2022,shimon_room_2022,bussandri_p1_2023,nir-arad_nitrogen_2023} and magic angle spinning conditions \cite{palani_dynamic_2024} showed enhancements exceeding 100 at several Tesla magnetic fields.
The DNP profiles at room temperature revealed a large number of narrow peaks ascribed to different hyperpolarization mechanisms including solid effect (SE), cross effect (CE) and truncated cross effect (tCE).


In this work, we study the hyperpolarization of \textsuperscript{13}C in diamond by DNP via P1 centers at 3.4\,T and 7\,T between 1.6\,K and 300\,K.
Nuclear polarization enhancements exceeding a factor of 100 for all employed conditions and polarization levels up to 38\% are found. 
The temperature-dependent changes of DNP profiles are complemented by longitudinal-detected (LOD) electron paramagnetic resonance (EPR) experiments.
In addition to the three P1 electron lines, a temperature-dependent broad electron line is detected.
The interplay between the different electron systems and their influence on DNP are discussed.

\section*{Methods}

\subsection*{Sample}
High-pressure high-temperature (HPHT) synthesized mono-crystalline diamonds with an average particle size of $10\pm2$\,\textmu m were purchased from Microdiamant AG (Switzerland).
In the Supplementary Material, three other diamond samples are characterized using EPR: (i) $<10$\,nm diamonds from Sigma-Aldrich (USA), (ii) nanodiamonds up to a size of 250\,nm from Microdiamant AG (Switzerland) and (iii) microdiamonds with $2 \pm 0.5$\,\textmu m average particle size as previously reported \cite{kwiatkowski_nanometer_2017}.
Diamonds were used without further treatment.

\subsection*{Dynamic nuclear polarization}
The DNP measurements were performed on home-built polarizers at 3.4\,T and 7\,T \cite{jahnig_dissolution_2017} with temperatures ranging from 1.6\,K to 300\,K.
The 3.4\,T (142\,MHz \textsuperscript{1}H Larmor frequency) polarizer was equipped with an OpenCore NMR spectrometer \cite{takeda_highly_2007,takeda_opencore_2008,takeda_chapter_2011} and the 7\,T (299\,MHz \textsuperscript{1}H Larmor frequency) with a Bruker Avance III console (Bruker BioSpin AG, Switzerland).
At 3.4\,T, a VDI (Virginia Diodes Inc., USA) microwave source with 400\,mW output power was coupled to a stainless steel waveguide.
At 7\,T, a VDI microwave source with 200\,mW output power was connected to an in-house electroplated low-loss, silver-coated stainless steel waveguide \cite{himmler_electroplated_2022}.
This permitted a similar microwave power in the sample space (estimated at 65\,mW for the 7\,T polarizer) for both set-ups.
Other details of the set-up are described elsewhere \cite{jahnig_dissolution_2017,himmler_electroplated_2022}. 

The absolute values of the polarization were calculated based on the average of two thermal equilibrium measurements (saturation, waiting time of approximately three times the nuclear spin-lattice relaxation time, followed by detection with a large flip angle pulse) at 3.4\,K.

\subsection*{Electron paramagnetic resonance}
EPR spectra at 3.4\,T or 7\,T were acquired with our in-house developed longitudinal-detection (LOD) EPR set-up \cite{himmler_electroplated_2022,himmler_improved_2023}. 
For LOD EPR measurements, a new coil and sample holder needed to be mounted on the cryostat inset while the MW set-up remains identical.
The VDI MW source was controlled through an attenuation voltage from a digital acquisition board (DAQ, National Instruments, USA) fed into the TTL input of the  MW source.
The EPR signal is detected with a home-built copper coil and the voltage was amplified before detection at 1\,MHz sampling rate with the same DAQ.

X-band EPR spectra were measured at room temperature with a Magnettech MiniScope MS5000 (Brucker Corp.).
The samples were filled in EPR tubes and placed at a controlled height within the spectrometer cavity which was automatically adjusted.
The spectra were taken with 30\,dB attenuation at a modulation amplitude of 0.2\,mT and a modulation frequency of 100\,kHz. 
The $g$-factor was verified and the number of spins was calculated using a standard TEMPO sample.
The uncertainty of measurements was estimated with separate technical replicates (multiple measurements of the same samples) of both TEMPO and a control porous Si sample \cite{von_witte_controlled_2024}.

\subsection*{Data analysis}
All data processing was performed with in-house developed MATLAB (MathWorks Inc., USA) scripts. 
All measurement uncertainties of the processed experimental data result from the 95\% fit intervals unless otherwise stated.
Experimental instabilities such as changes in the MW output power or minor temperature fluctuations as well as uncertainties in the thermal equilibrium measurement were considered negligible. 
In this work, we analysed the data in the time domain, fitted the FID with a combination of three oscillating exponentials (real part of the signal) and used the maximum of the fit (cf. Figs.~\ref{fig:SI_RawDataFit} and~\ref{fig:SI_BupAnalysis}, Supplementary Material). 
Among the possible models to fit the polarization build-ups and decays, we chose a stretched exponential function (cf. Eq.~\eqref{eq:StretchExp} and Figs.~\ref{fig:SI_BupStretchedExpParameters} and~\ref{fig:SI_BupPowerParameters}, Supplementary Material).
A comparison of the different analysis and fit methods can be found in Fig.~\ref{fig:SI_BupAnalysis}, Supplementary Material.
The Matlab scripts are available together with the experimental data (cf. Data Availability section).

\section*{Results} 

\subsection*{Dynamic nuclear polarization at different temperatures and fields}

\begin{figure}[t]
	\centering
	\includegraphics[width=\linewidth]{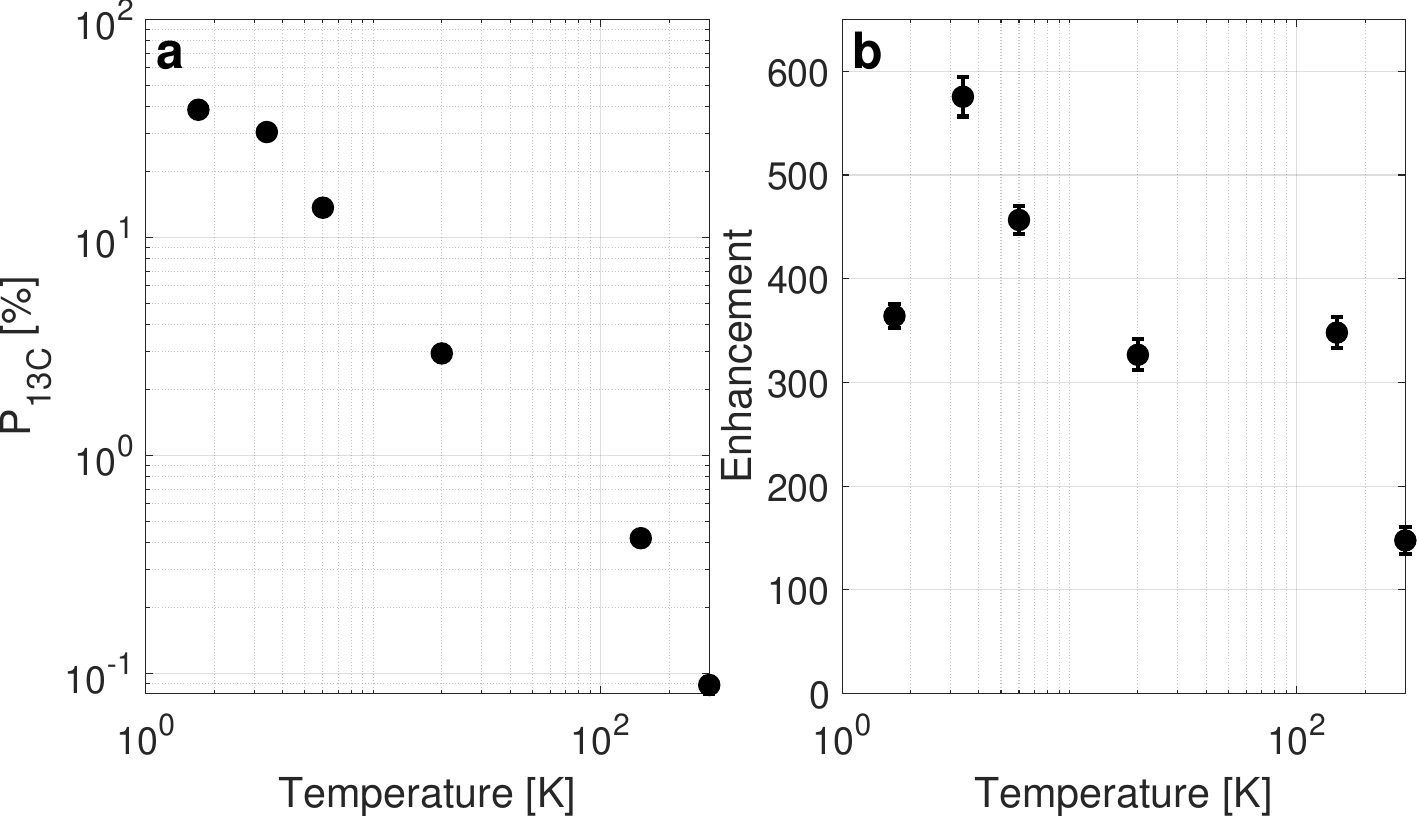}
	\caption{\textbf{(a)} Steady-state nuclear hyperpolarization levels and \textbf{(b)} enhancements between 1.7 and 300\,K at 7\,T and 196.830\,GHz.}
	\label{fig:Fig1}
\end{figure} 

\begin{figure*}[bht]
    \centering
	\includegraphics[width=\linewidth]{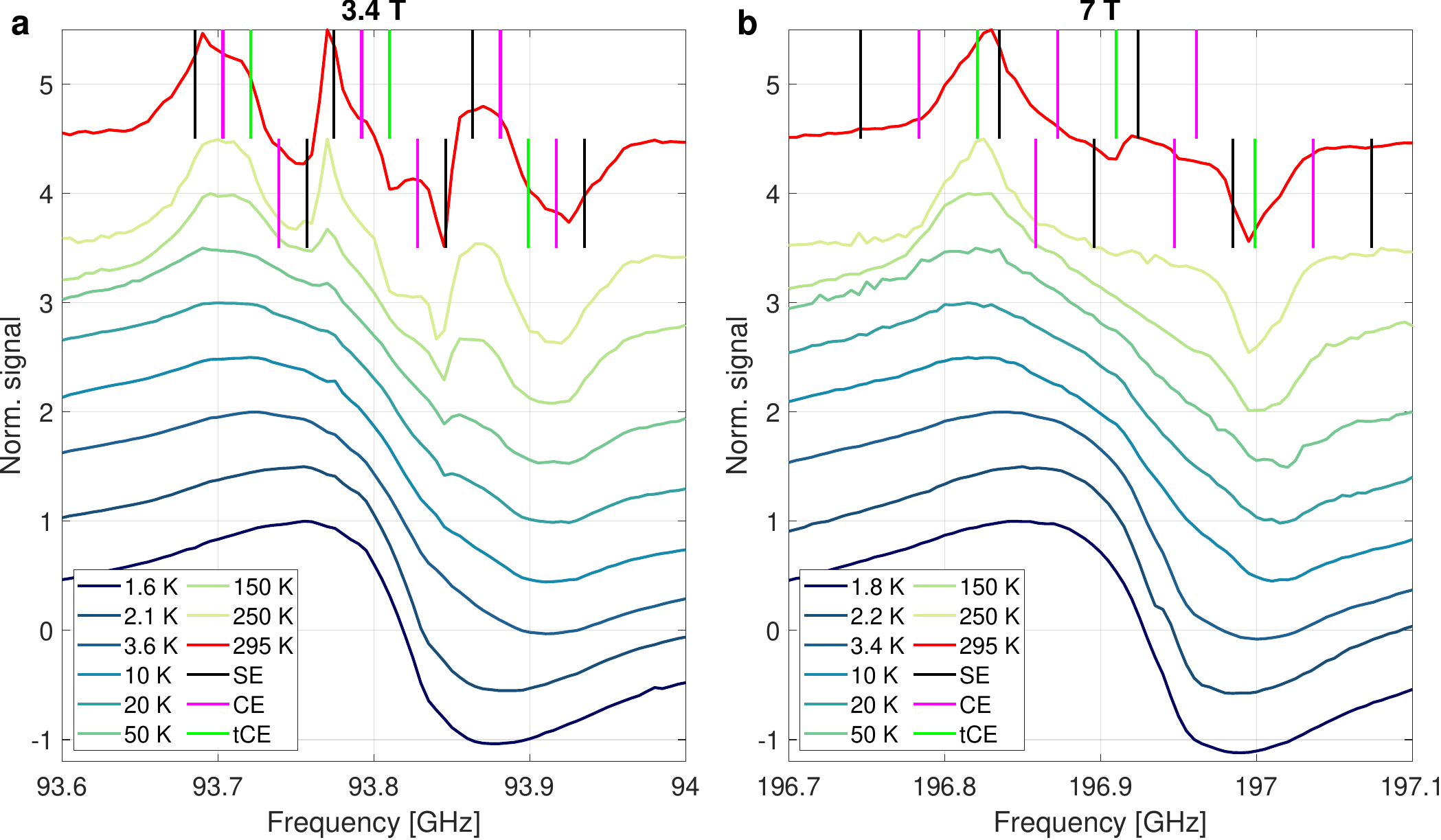}
	\caption{Selected DNP profiles between 295\,K and 1.6\,K for \textbf{(a)} 3.4\,T and \textbf{(b)}) 7\,T with frequencies of different DNP mechanism (SE: solid effect, CE: cross effect, tCE: truncated cross effect) indicated by vertical lines. 
    An extended data set of DNP profiles can be found in Fig.~\ref{fig:SI_DNPprofilesAll}, Supplementary Material.
    DNP profiles are vertically offset by 0.5 for clarity and twice the offset for 295\,K.}
	\label{fig:Fig2}
\end{figure*}

DNP was efficient at all temperatures between 1.7\,K and 300\,K at 3.4\,K and 7\,T as enhancements exceeding 100 for all temperatures were measured (cf. Fig.~\ref{fig:Fig1}b).
The observed room temperature polarization at 7\,T of 0.09\% (enhancement of 150) exceeds the achievable room temperature nuclear hyperpolarization with NV centers in diamond microparticles of around 0.04\% (enhancement of 1500 at 0.29\,T) \cite{blinder_13c_2024}.
At low temperature, the nuclear hyperpolarization is comparable to \textsuperscript{1}H spins in a glassy water-glycerol matrix doped with TEMPO \cite{von_witte_relaxation_2024} but lower than \textsuperscript{13}C-isotopically labelled pyruvic acid doped with trityl for which polarization levels up to 70\% have been reported \cite{macholl_trityl_2010,ardenkjaer-larsen_cryogen-free_2019}. 
Lowering the temperature from 300\,K to 3.4\,K results in an approximately exponential increase of the polarization (cf. Fig.~\ref{fig:Fig1}a).
Lowering the temperature from 3.4\,K to 1.7\,K increases the electron polarization from 88\% to 99\% and the nuclear polarization from 32\% to 38\%.
Together with the approximately exponential increase in polarization with decreasing temperature, this suggests that the nuclear hyperpolarization only depends on the thermal electron polarization and not on changes of electronic relaxation properties as often encountered with cooling.
Using notions of compartment modelling \cite{von_witte_modelling_2023, von_witte_relaxation_2024,von_witte_two-electron_2024}, the balance between DNP injection and relaxation rate constants appears to be independent of temperature. 

Fig.~\ref{fig:Fig2} compares the DNP profiles between 1.6\,K and 295\,K for 3.4\,T and 7\,T.
At room temperature (Fig.~\ref{fig:Fig2}a), several DNP peaks can be identified which can be linked to solid effect (SE), cross effect (CE) and truncated cross effect (tCE) DNP.

At 3.4\,T (Fig.~\ref{fig:Fig2}a), most of the features of the high temperatures profile disappear below 50\,K.
Below 50\,K, the DNP profiles show two broad DNP lobes with nearly symmetric intensities for spin-up and spin-down DNP.
At 7\,T (Fig.~\ref{fig:Fig2}b), the high temperature profile is dominated by two triangularly shaped peaks.
These peaks become smoother with decreasing temperatures.
Besides this, an additional smaller peak at the center of the DNP profile becomes visible at 295\,K but soon disappears with cooling.

\subsection*{Electronic spin system}

\begin{figure*}[t]
    \centering
	\includegraphics[width=\linewidth]{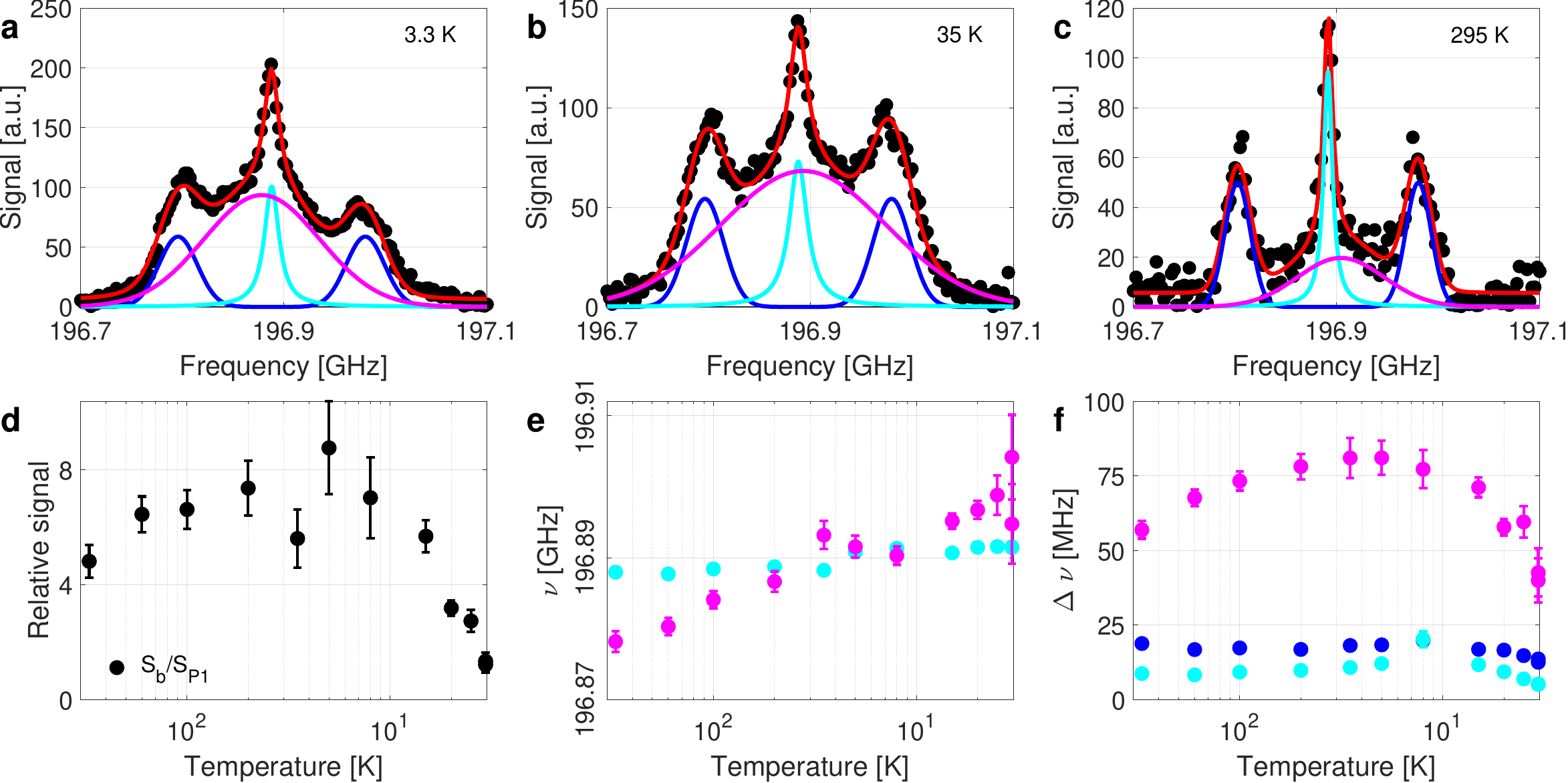}
	\caption{\textbf{(a-c)} Longitudinal-detected (LOD) electron paramagnetic resonance (EPR) profiles at 7\,T for different temperatures. 
    The LOD EPR spectra are fitted with Eq.~\eqref{eq:EPRmodel} as described in the main text.
    \textbf{(d)} Comparison of the signal amplitude of the broad component compared to the \textsuperscript{14}N hyperfine-split P1 center line. 
    \textbf{(e)} The fitted center frequencies of P1 center ($m_I = 0$, cyan) and the broad (magenta) component. 
    The broad component has a similar $g$-factor as the P1 center although a weak temperature dependence. 
    \textbf{(f)} Line widths of the Gaussian $m_I = \pm 1$ (blue), Lorentzian $m_I = 0$ (cyan) and broad Gaussian (magenta) lines. 
    Uncertainties can be smaller than the symbols for all fit parameters.
    All data was acquired at 7\,T and with full MW power. 
    Lower MW power causes a reduced signal (cf. Fig.~\ref{fig:SI_TorreyPower}, Supplementary Material) but results in qualitatively similar LOD profiles (data not shown but available, cf. Data Availability section).}
	\label{fig:Fig3}
\end{figure*} 

To better understand the observed changes in the DNP profiles of diamond, we performed longitudinal-detected (LOD) EPR measurements under DNP conditions (see Methods).
Due to improvements in our LOD EPR set-up \cite{himmler_electroplated_2022}, we were able to detect the diamond LOD EPR signal from 3.3\,K to 300\,K.
As evident in the EPR spectra shown in Fig.~\ref{fig:Fig3}a-c, the \textsuperscript{14}N hyperfine coupling to the P1 center is around 92\,MHz, which is in agreement with the literature values of $A_\parallel = 82$\,MHz and $A_\perp = 114$\,MHz \cite{smith_electron-spin_1959}.
However, the observed LOD EPR spectra cannot be explained based on three peaks originating from the \textsuperscript{14}N hyperfine split P1 electron system alone.
To fit the observed LOD EPR spectra, a combination of a Lorentzian line for the \textsuperscript{14}N $m_I = 0$, two Gaussians for the $m_I = \pm 1$ and an additional broader Gaussian line was assumed.
Specifically, we used
\begin{align}
    S_\mathrm{EPR} &= \frac{S_\mathrm{P1}}{\sqrt{2\pi}\sigma_{\pm 1}} \left[e^{-\frac{\left(\nu - (\nu_\mathrm{P1} - A_\mathrm{P1})\right)^2}{2\sigma_{ \pm 1}^2}} + e^{-\frac{\left(\nu - (\nu_\mathrm{P1} + A_\mathrm{P1})\right)^2}{2\sigma_{ \pm 1}^2}} \right] ...  \label{eq:EPRmodel} \nonumber \\
    &+ \frac{S_\mathrm{P1}}{\pi}\frac{\sigma_0}{\left(\nu - \nu_\mathrm{P1}\right)^2 + \sigma_0^2} 
    + \frac{S_\mathrm{b}}{\sqrt{2\pi}\sigma_\mathrm{b}} e^{\frac{\left(\nu-\nu_\mathrm{b}\right)^2}{2\sigma_\mathrm{b}^2}} + S_\mathrm{offset}
\end{align}
where 0 and $\pm 1$ subscripts indicate the different \textsuperscript{14}N hyperfine contributions of the P1 centers; the $\mathrm{b}$ subscript refers to a broad Gaussian line; $\sigma$ are the respective line widths; $S$ are the signal amplitudes of the different contributions; $\nu_\mathrm{P1}$ refers to the P1 center frequency of the $m_I = 0$ contribution and $A_\mathrm{P1}$ is the orientation averaged hyperfine coupling of the $m_I = \pm 1$ contributions.
The assumed model ensures that all \textsuperscript{14}N hyperfine contributions have the same intensity (area under the curve (AUC), total number of electrons) and the Gaussian line shape for the $m_I = \pm 1$ contributions approximates the powder broadening \cite{shimon_large_2022} of these lines due to the hyperfine coupling.
An example of the quality of the fit of the model applied to a measured EPR spectrum is shown in Fig.~\ref{fig:Fig3}a-c for LOD profiles at 3.3\,K, 35 \,K and 295\,K.

The fits reveal a large contribution of the broad component to the total LOD EPR signal at low and intermediate temperatures as shown in Fig.~\ref{fig:Fig3}d.
At 300\,K, the LOD EPR profile has only a weak broad component.
The resonance frequency of the broad component as given by its $g$-factor is similar to the resonance frequency of the P1 center.
In contrast to the P1 center, the center frequency of the broad component appears weakly temperature dependent relative to the frequency of the P1 center (cf. Fig.~\ref{fig:Fig3}e), which rationalizes the observed asymmetry of the EPR spectra at low temperature.
Moreover, the line width of the broad component shows a more pronounced temperature dependence (cf. Fig.~\ref{fig:Fig3}f) with the largest line widths at intermediate temperatures of tens of kelvin and the narrowest line at 300\,K.

\begin{figure*}[th]
	\centering
	\includegraphics[width=\linewidth]{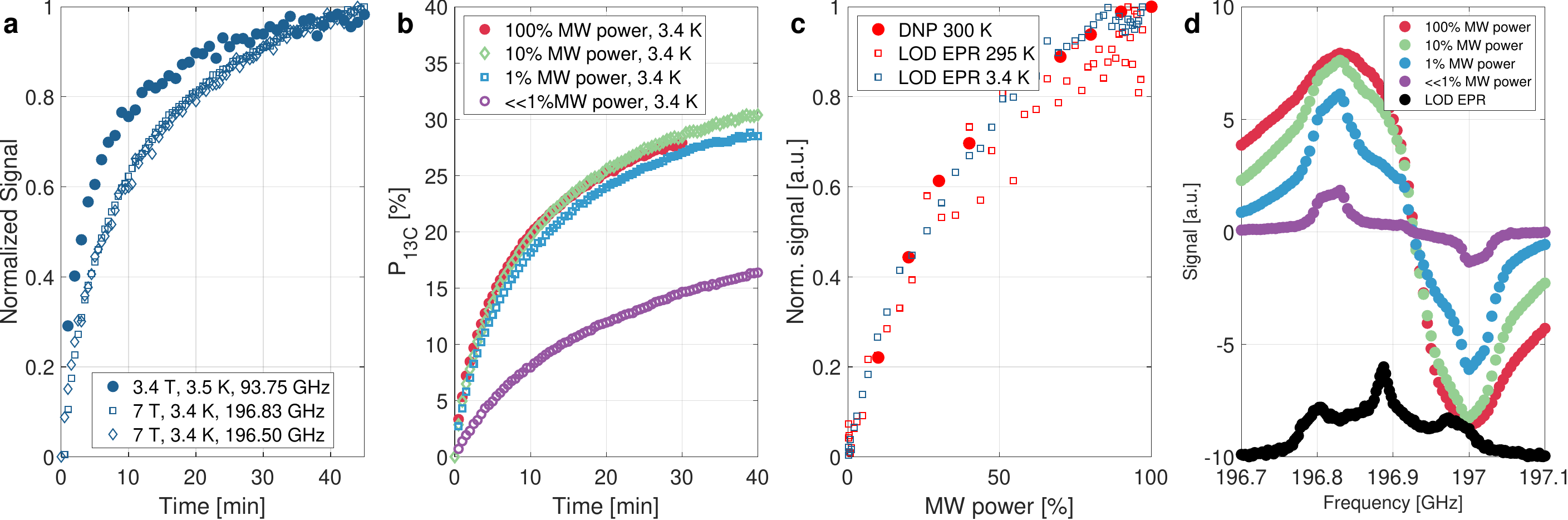}
	\caption{\textbf{(a)} Hyperpolarization build-up experiments around 3.4\,K for 3.4\,T and 7\,T. 
    At 7\,T, the build-up dynamics is independent of the MW frequency and in general slower than at 3.4\,T.
    At 7\,T, 3.4\,K and 196.50\,GHz, a nuclear polarization of 5.8$\pm$0.2\% is reached compared to 31$\pm 2$\% at 196.83\,GHz.
    \textbf{(b)} Power dependence of the build-up at 3.4\,K, 7\,T and 196.83\,GHz. 
    For the lowest power, the power is not exactly known but far below 1\% of the maximum MW power available.
    The fit parameters can be found in Fig.~\ref{fig:SI_BupPowerParameters}, Supplementary Material.
    \textbf{(c)} Power dependence of the DNP signal after 60\,s at 196.830\,GHz (7\,T) and 300\,K (filled symbols) and of the LOD EPR signal at 3.4\,K and 295\,K (open symbols).
    Each data set is normalized to its respective maximum measured signal.
    LOD EPR power curves between 3.4\,K and 295\,K are analyzed in Sec.~\ref{sec:SI_LOD} and in Fig.~\ref{fig:SI_TorreyPower} of the Supplementary Material.
    \textbf{(d)} DNP profiles for different MW output powers overlaid with the LOD EPR spectrum at 3.4\,K and 7\,T with 15\,s of DNP prior to detection.}
	\label{fig:Fig4}
\end{figure*}

\subsection*{Microwave frequency and power}

Fig.~\ref{fig:Fig4}a compares nuclear hyperpolarization build-up curves at 3.4\,K at 3.4\,T and 7\,T as well as curves at 7\,T for different microwave frequencies. 
While measurements at 7\,T in general show a slower polarization build-up, the polarization build-up appears independent of the MW frequency.
Specifically, we compare a DNP build-up at the MW frequency with the highest signal (196.83\,GHz) with a build-up at an frequency with a lower DNP enhancement at 7\,T (196.5\,GHz, cf. Fig.~\ref{fig:Fig2}b).

Figs.~\ref{fig:Fig4}b and c compare the power dependence of the DNP signal at 7\,T for 3.4\,K and 300\,K.
At 300\,K, the DNP signal shows a strong dependence on the MW power with the signal (cf. Fig.~\ref{fig:Fig4}c) increasing by more than fourfold if the MW power is increased from 10 to 100\% (around 20 to 200\,mW of output power and 6.5 to 65\,mW at the sample space, cf. Methods).
The nearly identical DNP build-up curves for 1, 10 and 100\% MW power at 3.4\,K (cf. Fig.~\ref{fig:Fig4}b) are in contrast to the pronounced power-dependence of the LOD EPR signal as displayed in Fig.~\ref{fig:Fig4}c.
At 295\,K, the LOD EPR and DNP signal both directly follow the saturation of the electron line (cf. Sec.~\ref{sec:SI_LOD}, Supplementary Material).
At 3.4\,K, the LOD EPR signal follows a similar trend as at 295\,K, while the DNP signal appears independent of the MW power for MW powers $\geq 1$\% and with this seemingly independent of the electron saturation (cf. Fig.~\ref{fig:Fig4}c).
Reducing the power to $\ll 1$\% MW power reduces the nuclear steady-state polarization achieved with DNP and leads to much longer build-up times (cf. Figs.~\ref{fig:Fig4}b and \ref{fig:SI_BupPowerParameters} Supplementary Material).
We emphasize that it is possible to achieve a nuclear hyperpolarization of 20\% for MW powers at the sample $\ll 1$\,mW.

Reducing the MW power changed the shape of the DNP profiles (cf. Fig.~\ref{fig:Fig4}d).
This is more prominent at $\leq 1$\% than at 10\% MW power, with narrow peaks at 196.83 and 197.00\,GHz combined with broader shoulders around the electron resonance frequency.
The frequency of 196.83\,GHz coincides with the frequencies of the highest DNP enhancements for higher MW powers and is identical to the maximum at 295\,K.
The 30-40\,MHz shift between the LOD and DNP profiles results from the temperature- and impurity-dependent diamangetic susceptibility of the copper \cite{bowers_magnetic_1956,schenck_role_1996} LOD EPR Helmholtz coil \cite{himmler_electroplated_2022}.

\section*{Discussion}

The room temperature DNP profiles at 3.4\,T and 7\,T shown in Fig.~\ref{fig:Fig2} are similar to those presented in Refs.~\cite{shimon_large_2022,bussandri_p1_2023}.
Nitrogen concentrations between 10 and 100\,ppm, 110-130\,ppm or less than 200\,ppm were reported for the samples used in Refs.~\cite{shimon_large_2022,bussandri_p1_2023}, while our sample contained around 54\,ppm of defects of which around 58\% are P1 centers (cf. Sec.~\ref{sec:SI_X-band}, Supplementary Material).
This suggests that the change of DNP and LOD EPR measurements with temperature reported herein appear representative for other diamond microparticles too. 

We emphasize that it is challenging to understand the DNP in diamond owing to a range of different defects, the interplay between these, possibly spatial inhomogeneity in terms of defect distribution, and different sample manufacturing procedures.
In the following, we will discuss observed temperature-dependent changes of DNP in diamond.

\subsection*{Polarization pathway}

The $10 \pm 2$\,\textmu m-sized diamond sample studied contained around 54\,ppm of defects of which 58\% were P1 centers (cf. Sec.~\ref{sec:SI_X-band}, Supplementary Material).
Assuming a homogeneous distribution of defects throughout the sample's bulk, the average distance between two unpaired electrons is estimated as $r_\mathrm{e-e} \approx n_\mathrm{e}^{-1/3} \approx 4.7$\,nm with $n_\mathrm{e}$ the electron concentration per unit volume.
The dipolar hyperfine coupling prefactor is $d_\mathrm{hfs} = \mu_0/8\pi^2 \cdot \hbar \gamma_\mathrm{e} \gamma_\mathrm{^{13}C}/(r_\mathrm{e-e}/2)^3 \approx 1.5$\,kHz for a nuclear spin at $r_{e-e}/2$ away from an electron (we ignore the other neighboring electrons for simplicity).
The $zz$-part of the hyperfine coupling describing the energetic shift of a nuclear spin has an additional prefactor of 2 and an angular dependence ($\left(3\cos^2{\theta} - 1\right)/2$ with $\theta$ denoting the angle between the two spins and the main magnetic field $B_0$).
Therefore, most nuclear spins will have a hyperfine coupling of a few kHz to an electron spin in their vicinity.

We note that the hyperfine coupling exceeds the zero-quantum (ZQ) line width $\Delta \nu_\mathrm{ZQ}$ as simulated from first principles ($\Delta \nu_\mathrm{ZQ} \approx 200-400$\,Hz, which corresponds to a nuclear spin diffusion coefficient of 20-40\,nm\textsuperscript{2}/s in a lattice approach or 4-8\,nm\textsuperscript{2}/s in a nearest neighbor approach \cite{von_witte_role_2024}).
We note that this estimate for the ZQ line width, which is similar to the experimentally accessible single-quantum (SQ) line width \cite{von_witte_role_2024}, is in good agreement with measured line widths in low defect diamonds of around 250\,Hz \cite{mark_henrichs_nuclear_1984}.
For spins with energy differences exceeding the ZQ line width, the probability of nuclear dipolar flip-flops, which are macroscopically considered as spin diffusion, vanishes.
Hence, spin diffusion is suppressed in the sample.
Owing to the small average nearest neighbor electronic and nuclear dipolar couplings of 0.5\,MHz and 100\,Hz, electron-nuclear four-spin flip-flops \cite{von_witte_two-electron_2024} do not lead to a significant nuclear spin diffusion either.
The absence of spin diffusion in diamond is in agreement with findings in Ref.~\cite{beatrez_electron_2023}.
The combination of few kHz hyperfine couplings and suppressed spin diffusion suggests that nuclei are preferentially hyperpolarized by a direct electron-nuclear polarization transfer.

Further evidence for a limited role of spin diffusion and a larger influence of direct hyperpolarization from electrons to nuclei comes from the stretched exponent of the fitted build-up curves (cf. Sec.~\ref{sec:SI_build-up}, Figs.~\ref{fig:SI_BupStretchedExpParameters}b and \ref{fig:SI_BupPowerParameters}c of the Supplementary Material) with most of the exponents around 0.8.
In Sec.~\ref{sec:SI_uncoupledCompartments}, Supplementary Material, a rate-equation model of hyperpolarization for infinitely-many uncoupled compartments (without spin diffusion, only hyperpolarization and relaxation by hyperfine coupling to the central electron with $r^{-3}$ scaling) is discussed.
The case of infinitely many uncoupled compartments describes the long-time behaviour of systems without spin diffusion and only direct DNP and relaxation through the electrons. 
For short time scales, systems with paramagnetic relaxation and without nuclear spin diffusion have been described with a stretched exponent of 0.5 \cite{blumberg_nuclear_1960, tse_nuclear_1968} while we find exponents close to 2/3 (cf. Sec.~\ref{sec:SI_uncoupledCompartments}, Supplementary Material).
For hyperpolarization with fast spin diffusion compared to the hyperpolarization injection ($k_\mathrm{W}$) and relaxation ($k_\mathrm{R}$) rate constants such that the build-up time constant is given by $\tau_\mathrm{bup} = \left(k_\mathrm{W} + k_\mathrm{R} \right)^{-1}$, a mono-exponential build-up is found \cite{von_witte_modelling_2023}.
Hence, the stretch exponent of around 0.8 might be interpreted as direct hyperpolarization being the main polarization pathway while spin diffusion plays a minor role.

\subsection*{Electronic spin system}

\begin{figure}[th]
	\centering
	\includegraphics[width=\linewidth]{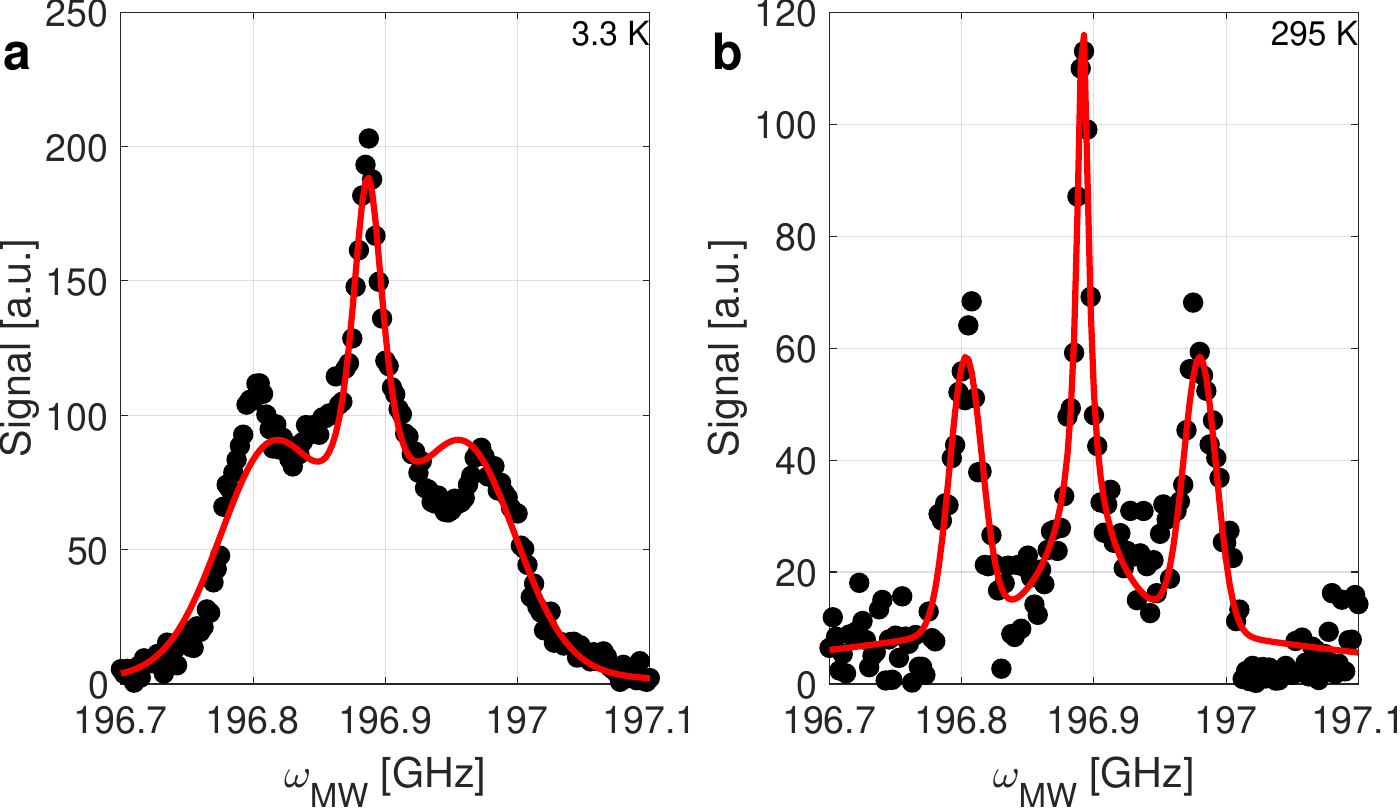}
	\caption{ Longitudinal-detected (LOD) electron paramagnetic resonance (EPR) profiles at 7\,T and \textbf{(a)} 3.3\,K or \textbf{(b)} 295\,K. 
    The LOD EPR spectra are fitted with Eq.~\eqref{eq:EPRmodel_expandedP1} as described in the main text.}
	\label{fig:Fig5}
\end{figure}

The LOD EPR profiles were fitted with a combination of powder broadened P1 centers and a broad (spin-1/2) defect (cf. Fig.~\ref{fig:Fig3}, Eq.~\eqref{eq:EPRmodel} and Sec.~\ref{sec:SI_LOD}, Supplementary Material).
Inspired by Refs.~\cite{bussandri_p1_2023,nir-arad_nitrogen_2023,palani_dynamic_2024}, the measured LOD EPR profiles were fitted with a combination of narrow and broad P1 centers.
The narrower of the two P1 populations is supposed to describe rather isolated P1 centers while the second broader P1 population is associated with cluster-broadened P1 centers.
The fit function for this is
\begin{align}
    S_\mathrm{EPR} &= \frac{S_{\mathrm{P1}, n}}{\sqrt{2\pi}\sigma_{\pm 1, n}} \left[e^{-\frac{\left(\nu - (\nu_\mathrm{P1} - A_\mathrm{P1})\right)^2}{2\sigma_{ \pm 1, n}^2}} 
    + e^{-\frac{\left(\nu - (\nu_\mathrm{P1} + A_\mathrm{P1})\right)^2}{2\sigma_{ \pm 1, n}^2}} \right] \nonumber \\ 
    &+ \frac{S_{\mathrm{P1}, n}}{\pi}\frac{\sigma_{0, n}}{\left(\nu - \nu_\mathrm{P1}\right)^2 + \sigma_{0, n}^2} \nonumber \\ 
    &+ \frac{S_{\mathrm{P1}, b}}{\sqrt{2\pi}\sigma_{\pm 1, b}} \left[e^{-\frac{\left(\nu - (\nu_\mathrm{P1} - A_\mathrm{P1})\right)^2}{2\sigma_{ \pm 1, b}^2}} 
    + e^{-\frac{\left(\nu - (\nu_\mathrm{P1} + A_\mathrm{P1})\right)^2}{2\sigma_{ \pm 1, b}^2}} \right] \nonumber \\ 
    &+ \frac{S_{\mathrm{P1}, b}}{\pi}\frac{\sigma_{0, b}}{\left(\nu - \nu_\mathrm{P1}\right)^2 + \sigma_{0, b}^2} 
    + S_\mathrm{offset} \label{eq:EPRmodel_expandedP1} 
\end{align}
with the $n$ and $b$ subscripts referring to narrow (isolated) and broad (clustered) P1 contributions.  
The central P1 frequency ($g$-factor) and hyperfine coupling was assumed to be identical for the two P1 populations.
The fits of Eq.~\eqref{eq:EPRmodel_expandedP1} to the 7\,T LOD EPR spectra at 3.3\,K and 295\,K are shown in Fig.~\ref{fig:Fig5} with the remaining LOD EPR fits shown in Fig.~\ref{fig:SI_LODprofilesExpandedFitFunction}, Supplementary Material, and the fit parameters summarized in Fig.~\ref{fig:SI_LODprofilesExpandedFitFunctionFitParameters}, Supplementary Material.
At high temperatures, this two P1 population fitting model works quite well although it is rather insensitive to several fit parameters, e.g. relative signal between narrow and broad components and some of the broadening as displayed in Fig.~\ref{fig:SI_LODprofilesExpandedFitFunctionFitParameters}b and d, Supplementary Material.
At temperatures below around 200\,K as exemplified by Fig.~\ref{fig:Fig5}a for 3.3\,K, the two P1 population model struggles with fitting the peak frequencies (hyperfine coupling), peak heights, flanks and asymmetry of the LOD EPR profiles.
Fundamentally, this model is not capable of describing the asymmetry observed in the LOD EPR profiles as evident through asymmetric heights between the $m_I = -1$ and $m_I = +1$ lines - maybe most pronounced at 3.3\,K (cf. Fig.~\ref{fig:Fig5}a).

Therefore, we propose a fit model based on a single set of P1 centers and a broad (spin-1/2) defect which describes our sample better (cf. Figs.~\ref{fig:Fig3},~\ref{fig:SI_LODprofiles}, Supplementary Material, and Eq.~\ref{eq:EPRmodel}).
Owing to the large particle size of $10 \pm 2$\,\textmu m with its low surface-to-volume ratio compared to nanoparticles, it is unlikely that the additional broad line, which contributes around 42\% of the total defects in the sample (cf. Sec.~\ref{sec:SI_X-band}, Supplementary Material), arises from surface dangling bonds but is considered an additional bulk defect.
Different defects have been reported in diamonds with a larger number of nitrogen-based defects, which have similar $g$-factors as the P1 center \cite{loubser_electron_1978}.
In Sec.~\ref{sec:SI_defectDiscussion}, Supplementary Material, a selected group of nitrogen bulk defects is discussed.
We highlight here that so-called N2 and N3 centers could explain the observed broad line and the likely shortening of P1 electronic relaxation times as discussed in the next section.

\subsection*{Electronic and nuclear relaxation times}

The total electron concentration of 54\,ppm corresponds to around 200 \textsuperscript{13}C nuclei per electron for \textsuperscript{13}C at 1.1\% natural abundance in diamond.
Considering only the P1 centers, our sample contained around 350 \textsuperscript{13}C nuclei per P1 center.

In Ref.~\cite{reynhardt_temperature_1998}, diamond samples with 25 or 95\,ppm of P1 centers and without other (nitrogen) defects were investigated.
Accordingly, electronic $T_\mathrm{1,e}$ relaxation times of tens to hundreds of seconds around 10\,K with increasing relaxation times upon decreasing temperatures were found.
For such long electronic relaxation times, it is difficult to envision how the stretched DNP build-up times of 12\,min (cf. Fig.~\ref{fig:SI_BupStretchedExpParameters}a, Supplementary Material) with up to 38\% nuclear polarization at liquid-helium temperatures is feasible from such slow relaxing P1 centers.

Refs.~\cite{reynhardt_temperature_1998, reynhardt_spin_2003} reported samples containing N2 and N3 centers and found a shortening of the P1 and nuclear relaxation times, suggesting an interplay between the different bulk spin defects.
Shortened electron relaxation times of the P1 centers appear consistent with cryogenic X- and Q-band EPR experiments of diamond micro- and nanoparticles designed for hyperpolarized DNP  \cite{rej_hyperpolarized_2015, waddington_phase-encoded_2019,boele_tailored_2020}.
Further evidence for the shortening of P1 electron relaxation times comes from our LOD experiments with LOD time constants of a few hundred \textmu s (cf. Sec.~\ref{sec:SI_LOD} and specifically Fig.~\ref{fig:SI_electronTimes}, Supplementary Material).
The measured LOD time constants are comparable to other DNP radicals/ defects \cite{himmler_electroplated_2022}.


\subsection*{Microwave power dependence}

At 3.4\,K and 7\,T, the DNP build-up curves at the frequency with the highest enhancements were independent of MW powers exceeding 1\% (cf. Fig.~\ref{fig:Fig4}b), while the LOD EPR signal under the same conditions showed a power dependence (cf. Fig.~\ref{fig:Fig4}c and \ref{fig:SI_TorreyPower}, Supplementary Material).
In contrast, at 300\,K the DNP signal followed the electron saturation as measured in the LOD EPR power curve (cf. Fig.~\ref{fig:Fig4}c).

For TEMPO in \textsuperscript{1}H glassy matrices \cite{von_witte_relaxation_2024} at liquid-helium temperatures, a nearly MW power-independent DNP signal was accompanied by a decrease in build-up time for high MW powers.
This was attributed to an increase in triple spin flips as in cross effect (CE) DNP, which causes, on the one hand, DNP and, on the other hand, paramagnetic relaxation.
However, in the current case, the build-up times for the 1, 10 and 100\% MW power measurements were nearly identical (cf. Fig~\ref{fig:Fig4}b and Fig.~\ref{fig:SI_BupPowerParameters}b of the Supplementary Material), suggesting a different origin of the MW power-independent signal and with that a possible lack of relevance of triple spin flips.

Far below 1\% MW power (MW source not calibrated in this regime), the achievable steady-state polarization decreases and the build-up time is prolonged. 
For such low MW powers, the rather efficient DNP might be due to isolated defects with very long electronic relaxation times such that even a weak MW field causes electron saturation and enables an efficient DNP creation.
This qualitative difference in DNP generation is supported by the changes in the DNP profile with MW power (cf. Fig.~\ref{fig:Fig4}d).
The DNP profiles seem to consist of a broad and narrow component.
The broad component is dependent on the MW power, while the narrow component is mostly independent of MW power.
This could be explained by two different DNP processes with the broad component depending on the broad electron line while the narrow component depends on the $m_I = \pm 1$ P1 electron lines.
The narrow components resemble the shape of powder broadened  $m_I = \pm 1$ P1 electron lines (cf. Fig.~\ref{fig:SI_EPR_Konstantin_fits}, Supplementary Material and Ref.~\cite{shimon_large_2022}).

\subsection*{Summary}

In \textmu m-sized diamonds, DNP enhancements of several hundreds between 1.7\,K and 300\,K and few Tesla magnetic field are achievable with $\leq 200$\,mW of microwave power.
When lowering the temperature, the DNP profiles change from feature-rich to broad DNP lobes (positive and negative enhancements), indicative of different DNP origins. 
Our results suggest that P1 centers and a broad spin-1/2 electron line, tentatively associated with N2 or N3 centers, cause the observed DNP with nuclei hyperpolarized directly via hyperfine coupling rather than through suppressed spin diffusion. 
The interplay between different temperature-dependent electron systems may offer new possibilities to study dynamic nuclear polarization.



\nocite{terblanche_room-temperature_2000,terblanche_13c_2001,stoll_easyspin_2006,loubser_singly_1973,wyk_endor_1992,jardon-alvarez_enabling_2020}

\bibliography{notes,references}

\bibliographystyle{sciencemag}

\textbf{Acknowledgments:} GvW thanks Leon Rückert, Daphna Shimon, Chandrasekhar Ramanathan, Orit Nir-Arad, Ilia Kaminker and Tom Wenckebach for discussions.

\textbf{Funding:} ME acknowledges support by the Schweizerischer Nationalfonds zur Förderung der Wissenschaftlichen Forschung (grant no. 200020\_188988 and 200020\_219375).
KT and JOM acknowledge support by Research Council of Finland (grant no. 331371, 338733, and Flagship of Advanced Mathematics for Sensing Imaging and Modelling grant 358944), Finnish Cultural Foundation (North Savo regional fund) and Saastamoinen Foundation.
Financial support of the Horizon 2020 FETFLAG MetaboliQs grant is gratefully acknowledged.

\textbf{Author Contributions:} GvW, ME and SK conceptualized the research.
GvW performed DNP and LOD EPR experiments with the help from AH.
GvW analyzed the DNP and LOD EPR measurements.
KT performed and analyzed the X-band EPR measurements.
GvW prepared the original draft with contributions from KT. 
ME and SK acquired funding and provided supervision. 
JOM, ME and SK provided resources.
All authors reviewed and edited the draft..

\textbf{Competing Interests:} The authors declare that they have no competing interests.

\textbf{Data and Materials Availability:} All data needed to evaluate the conclusions in the paper are present in the paper and/or the Supplementary Materials.
The raw experimental data and the Matlab scripts for processing can be found under \url{https://doi.org/10.3929/ethz-b-000709870}.

\clearpage

\section*{Supplementary Materials: Temperature-dependent dynamic nuclear polarization of diamond }

\author
{Gevin von Witte$^{1,2}$, Aaron Himmler$^{2}$, Konstantin Tamarov$^{3}$, Jani O. Moilanen$^{4}$, Matthias Ernst$^{2}$,  Sebastian Kozerke$^{1\ast}$\\
\\
\normalsize{$^{1}$Institute for Biomedical Engineering, University and ETH Zurich, 8092 Zurich, Switzerland}\\
\normalsize{$^{2}$Institute of Molecular Physical Science, ETH Zurich, 8093 Zurich, Switzerland}\\
\normalsize{$^{3}$Department of Technical Physics, University of Eastern Finland, 70210 Kuopio, Finland}\\
\normalsize{$^{4}$Department of Chemistry, Nanoscience Center, 
University of Jyväskylä, Jyväskylä, Finland}\\
\\
\normalsize{$^\ast$To whom correspondence should be addressed; E-mail: kozerke@biomed.ee.ethz.ch}
}


\setcounter{equation}{0}
\setcounter{figure}{0}
\setcounter{table}{0}
\setcounter{section}{0}
\setcounter{page}{1}
\makeatletter
\renewcommand{\theequation}{S\arabic{equation}}
\renewcommand{\thefigure}{S\arabic{figure}}
\renewcommand{\thetable}{S\arabic{table}}
\renewcommand{\thesection}{S\arabic{section}}

\tableofcontents

\section{Build-ups and decay} \label{sec:SI_build-up}

\begin{figure*}[ht]
	\centering
	\includegraphics[width=\linewidth]{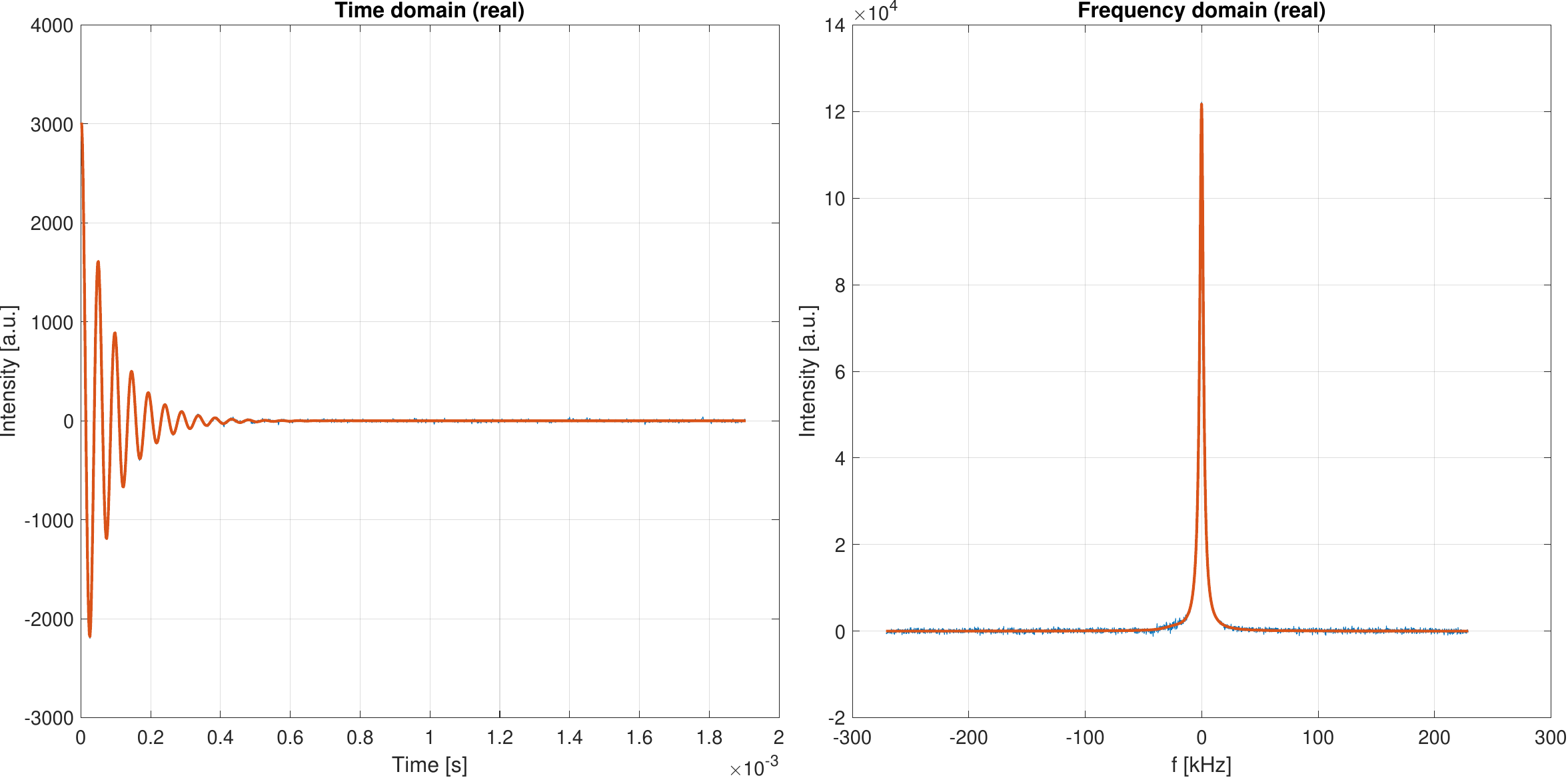}
	\caption{Experimental raw data in the frequency and time domain together with the respective fits.
    The displayed data belongs to the build-up shown in Fig.~\ref{fig:SI_BupAnalysis} after 10\,min of DNP.}
	\label{fig:SI_RawDataFit}
\end{figure*}

\begin{figure*}[ht]
	\centering
	\includegraphics[width=\linewidth]{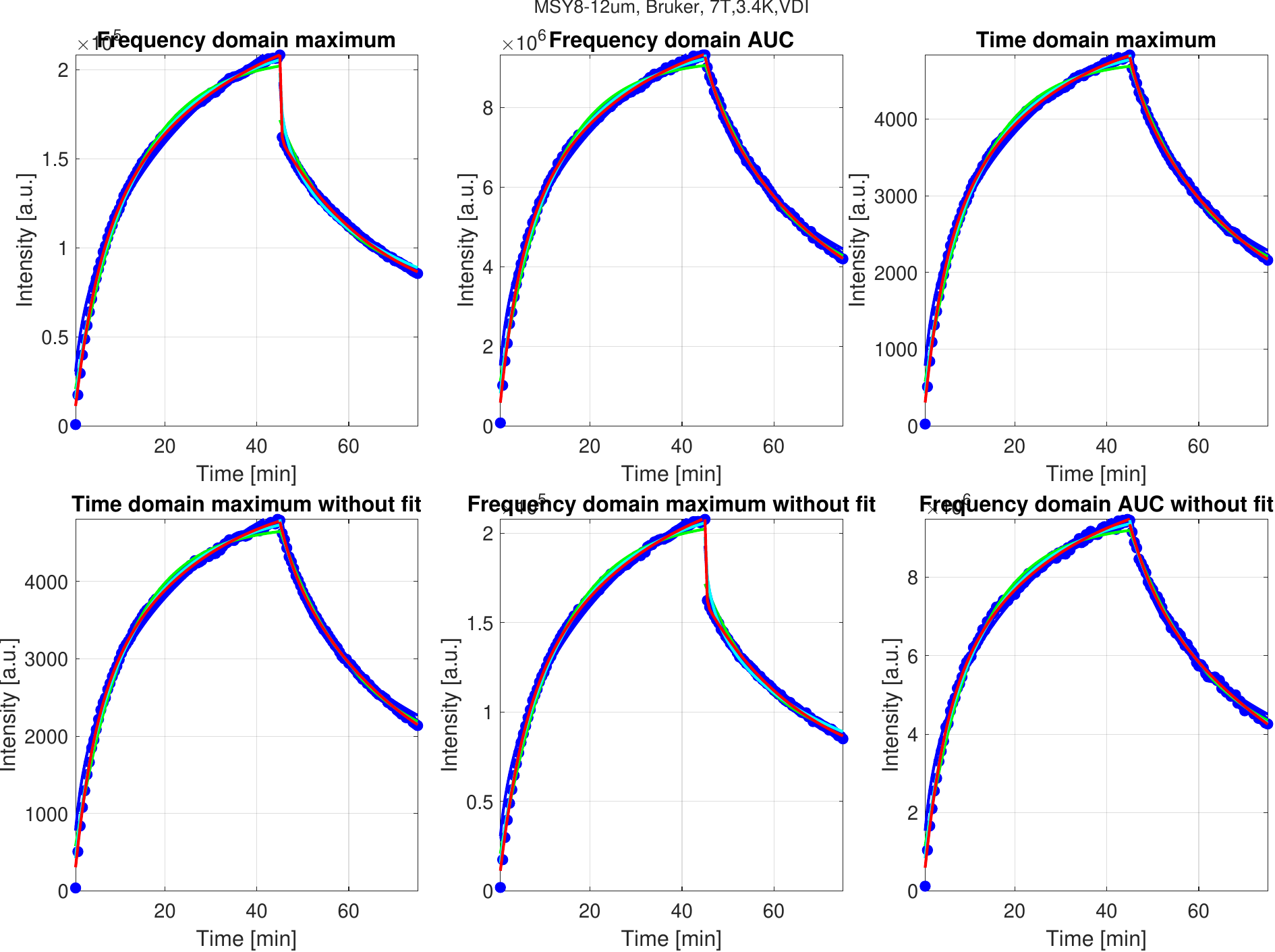}
	\caption{Different analysis methods for the build-up and decay (time or frequency domain, with or without fit, maximum or area under the curve(AUC)) as well as mono- (green), square-root- (a stretched exponential with $\kappa = 1/2$, blue), stretched (cyan) and bi-exponential (red) fits.}
	\label{fig:SI_BupAnalysis}
\end{figure*}

The measured free induction decays (FIDs) can be fitted either in the time domain or Fourier-transformed frequency domain as shown in Fig.~\ref{fig:SI_RawDataFit}.
In the frequency domain, a combination of three pseudo-Voigt functions is used to fit the spectrum. 
Subsequently, the signal for a given measurement can be evaluated as the area-under-the-curve (AUC) or the maximum.
In the time domain, a combination of three oscillating decaying exponential functions is used for the fits and the maximum fitted signal used.
Due to the high SNR in the presented measurements, fitting could be omitted and signals evaluated directly in the time or frequency domain.
All these analysis methods applied to a build-up and decay measurements are shown in Fig.~\ref{fig:SI_BupAnalysis}.
As the line widths in the sample change upon interruption of the MW irradiation (cf. Fig.~\ref{fig:SI_BupAnalysis} for frequency domain maximum), the frequency domain maximum is a poor choice as it is not invariant to line width changes.
We note that this change of signal is qualitatively different than the observed change in DNP signal for TEMPO in \textsuperscript{1}H glassy matrices \cite{von_witte_relaxation_2024} where the signal due to a change in the number of spins detectable through detection pulses (decoupling of electron-nuclear hyperfine coupling by MW irradiation).
In the current case, the line width of the detected NMR signal changes while the time domain maximum signal or integrated frequency domain signal are not affected by the change in MW irradiation.
Fitting in the frequency or time domain has the advantage of filtering out noise to some degree which is especially important for low experimental SNRs as encountered at high temperatures or at the outside of the DNP profiles.
Fitting in the time domain and extracting the signal from the maximum of the fit (zero FID time) appears as the best choice as it combines filtering of noise if the signal is low, e.g. at high temperatures or at the very outside of the DNP profile, with invariance to a change in line width.

The obtained build-up and decay curves are fitted with either a mono- or bi-exponential function, a square-root exponential (a stretched exponential with $\kappa = 1/2$) or a stretched exponential of the form
\begin{align}
	P = P_{0,s}\cdot \left(1-e^{-\left(t/\tau_\mathrm{s}\right)^\kappa}\right) \label{eq:StretchExp}
\end{align}  
with $P_{0,s}$ being the steady-state polarization of the stretched exponential build-up, $\tau_\mathrm{s}$ the build-up time constant and $\kappa$ is the stretch exponent.
Eq.~\eqref{eq:StretchExp} is for the build-up and the decay model is obtained by dropping the $1-$ part.
All these build-up and decay fits are shown in Fig.~\ref{fig:SI_BupAnalysis}.

\begin{figure*}[ht]
	\centering
	\includegraphics[width=0.8\linewidth]{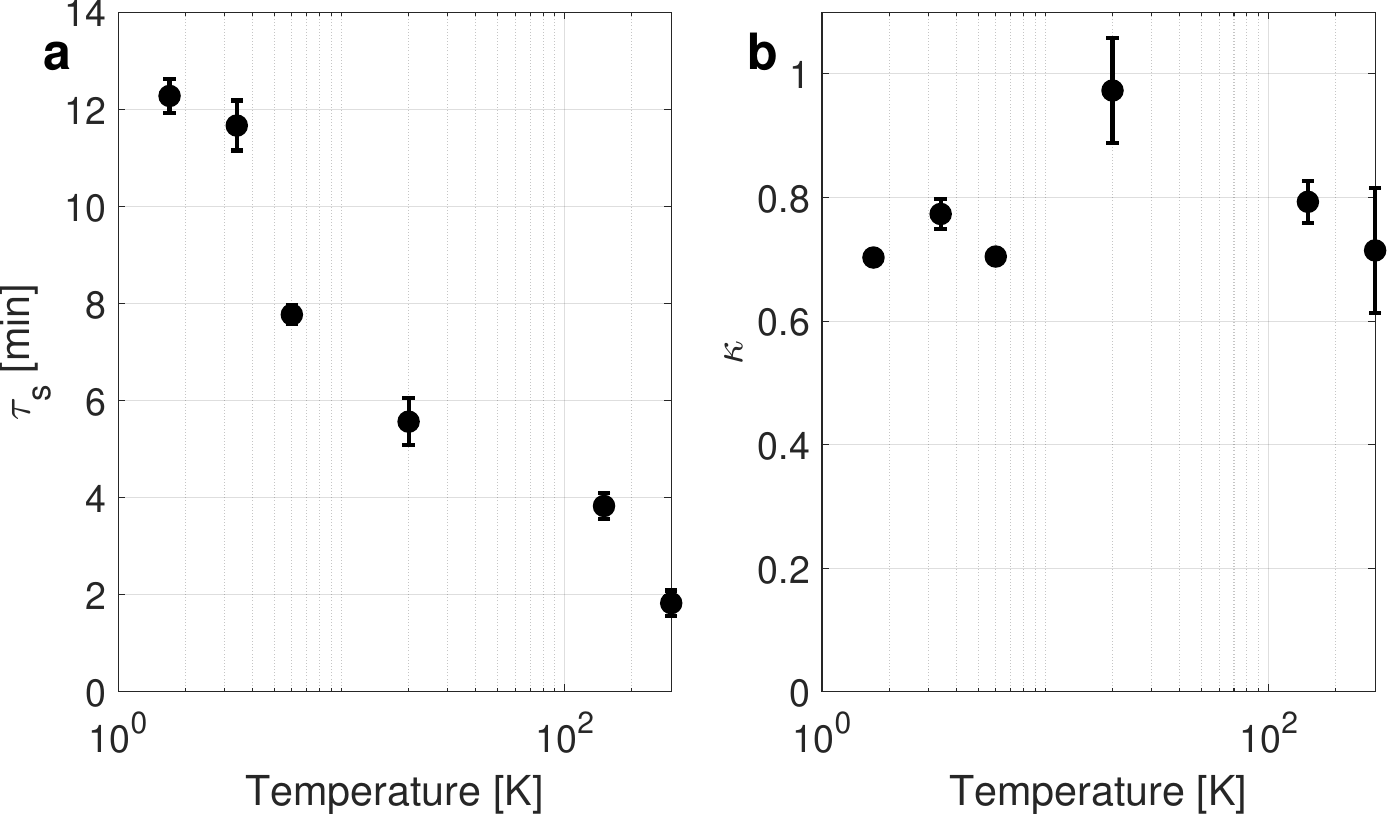}
	\caption{Stretched exponential build-up fit parameters \textbf{(a)} $\tau_\mathrm{s}$ and \textbf{(b)} $\kappa$ (cf. Eq.~\eqref{eq:StretchExp}). 
    The steady-state polarization for all the temperatures are given in Fig.~\ref{fig:Fig1}.
    Uncertainties might be smaller than symbols.}
	\label{fig:SI_BupStretchedExpParameters}
\end{figure*}

The stretched and bi-exponential gave the best results throughout this work. 
We chose a stretched exponential due to its more robust fitted parameters which are given in Figs.~\ref{fig:Fig1} and \ref{fig:SI_BupStretchedExpParameters}.
Furthermore, the stretched exponential ansatz has a direct interpretation in terms of a direct DNP transfer (cf. Sec.~\ref{sec:SI_uncoupledCompartments}) while for the bi-exponential ansatz could be interpreted as spin diffusion between different compartments causing different time scales.
The latter is hard to imagine with the high and rather homogeneous distribution of defects in the bulk as well as the estimated fast (with respect to the distances between defects) spin diffusion.

\begin{figure*}[ht]
	\centering
	\includegraphics[width=0.8\linewidth]{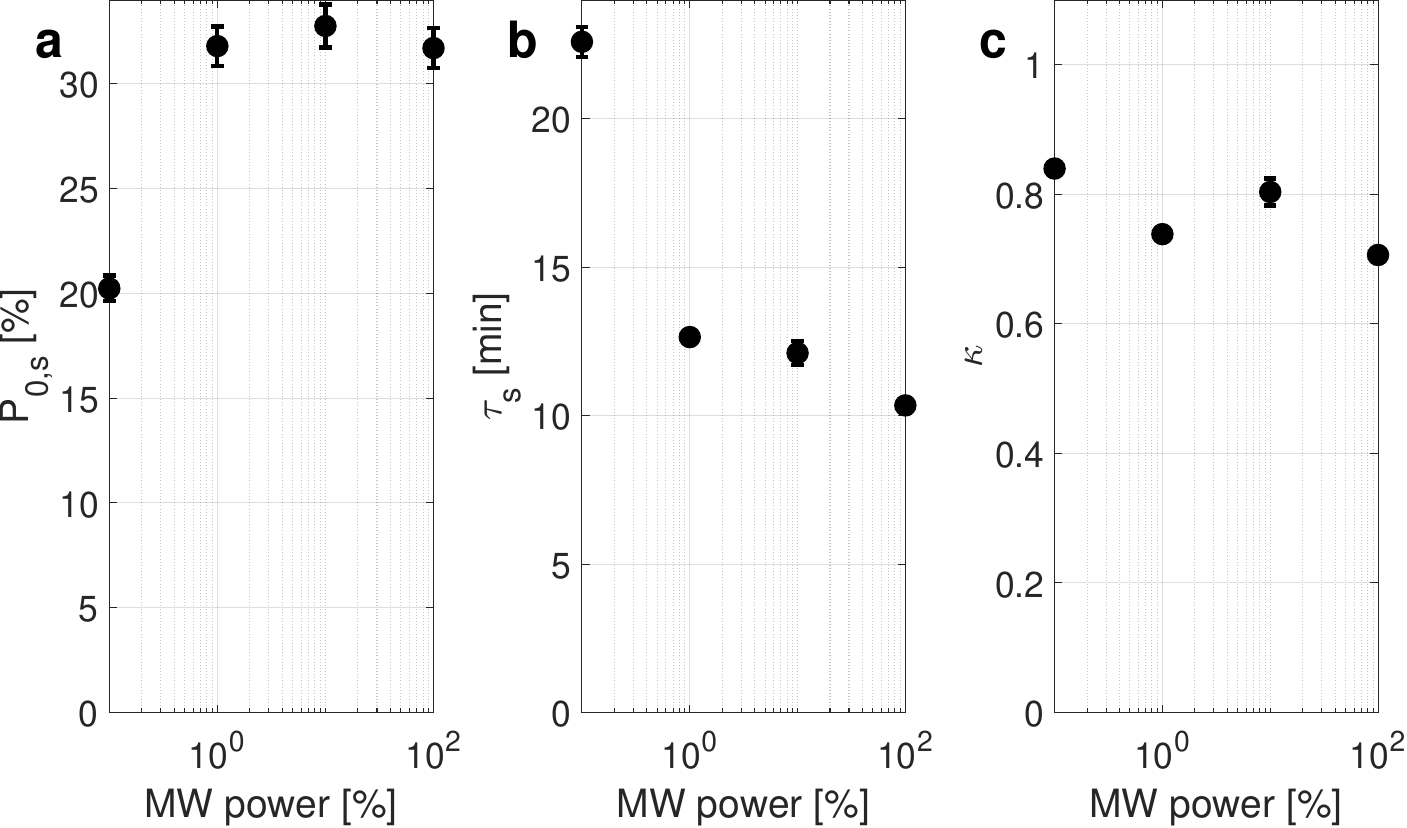}
	\caption{Fit parameters of the data shown in Fig.~\ref{fig:Fig4}c.
    The build-ups were fit with a stretched exponential ansatz (cf. Eq.~\eqref{eq:StretchExp}).
    The exact MW power for the lowest MW power is unknown but $\ll1$\%.}
	\label{fig:SI_BupPowerParameters}
\end{figure*}

For the hyperpolarization decay/ relaxation after the build-up experiment at 7\,T and 3.4\,K (cf. Fig.~\ref{fig:Fig4}a) as shown in Figs.~\ref{fig:Fig4} and \ref{fig:SI_BupAnalysis}, we find a decay time constant $\tau_\mathrm{s}^\mathrm{dec} = 27 \pm 4$\,min and a stretch exponent of $\kappa = 0.79 \pm 0.03$.
The stretch exponent is in good agreement with the build-up (cf. Figs.~\ref{fig:SI_BupStretchedExpParameters}b and \ref{fig:SI_BupPowerParameters}c).
The stretched exponential decay time constant is nearly twice as long as for the build-up which is attributed to the the vanishing DNP injection with the MW being switched off.
For more information on this, the interested reader is referred to Sec.~\ref{sec:SI_uncoupledCompartments}.

At 3.4\,T and 3.5\,K, the stretch exponent for the build-up and decay are $0.59 \pm 0.12$ and $0.71 \pm 0.03$. 
The corresponding stretched exponential time constants for the build-up and decay $6.3 \pm 1.5$ and $6.9 \pm 0.3$\,min.

\clearpage 


\section{DNP profiles} \label{sec:SI_DNPprofiles}

\begin{figure*}[ht]
    \centering
	\includegraphics[width=\linewidth]{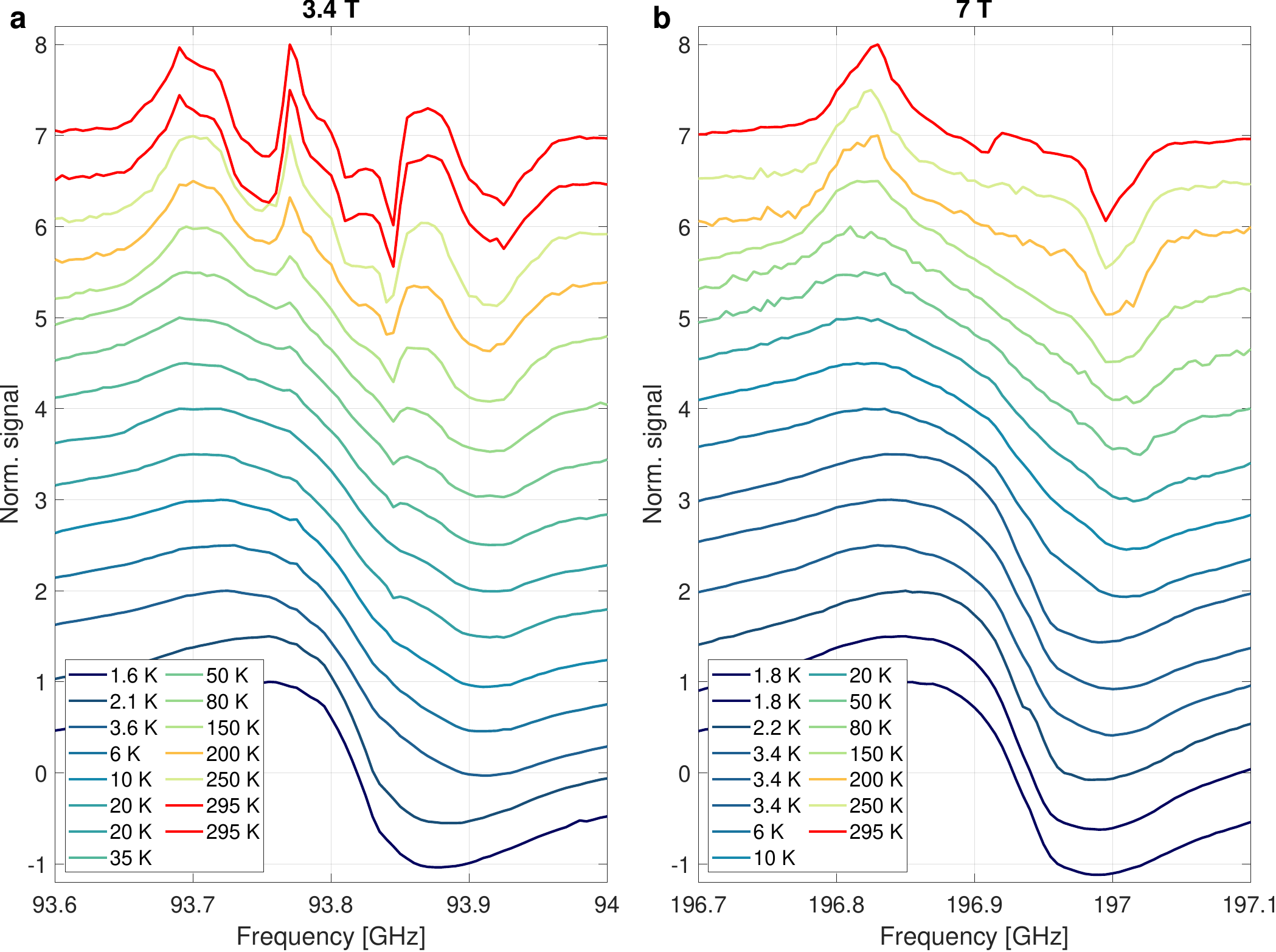}
	\caption{DNP profiles between 295\,K and 1.6\,K for \textbf{(a)} 3.4\,T and \textbf{(b)}) 7\,T. 
    DNP profiles are vertically offset by 0.5 for clarity.}
	\label{fig:SI_DNPprofilesAll}
\end{figure*}

Fig.~\ref{fig:SI_DNPprofilesAll} shows the full set of recorded DNP profiles at different temperatures and at 3.4\,T and 7\,T (cf. Fig.~\ref{fig:Fig2}).

\begin{figure*}[ht]
    \centering
	\includegraphics[width=\linewidth]{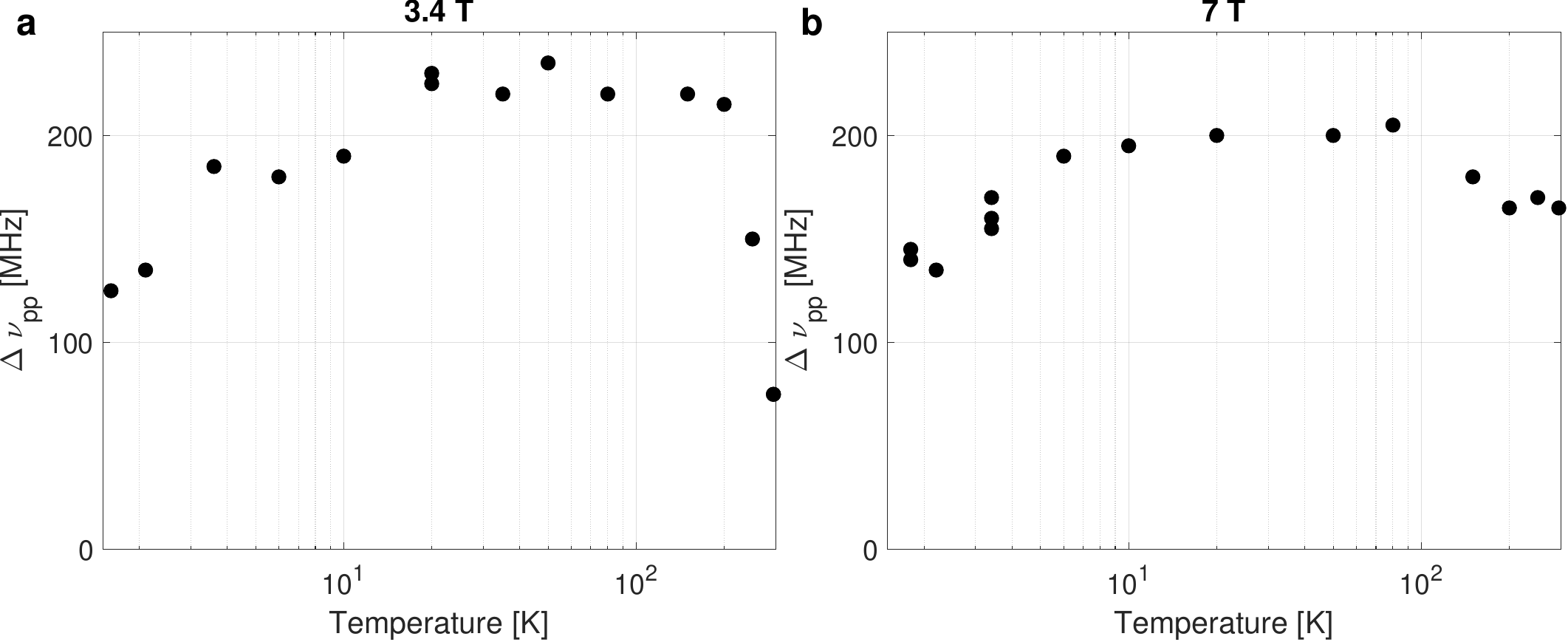}
	\caption{Frequency difference between the maximum and minimum of the DNP profiles ($\Delta \nu_\mathrm{pp}$) depending on the temperature for \textbf{(a)} 3.4\,T and \textbf{(b)} 7\,T.
    The corresponding DNP profles are show in Fig.~\ref{fig:SI_DNPprofilesAll}.}
	\label{fig:SI_DNPprofilesPeakToPeak}
\end{figure*}

The frequency difference between the DNP maximum and minimum $\Delta \nu_\mathrm{pp}$ are shown in Fig.~\ref{fig:SI_DNPprofilesPeakToPeak} for all DNP profiles from Fig.~\ref{fig:SI_DNPprofilesAll}.
At 3.4\,T (Fig.~\ref{fig:Fig2}a), the initial increase of $\Delta \nu_\mathrm{pp}$ is explained through the different temperature dependences of the two large inner DNP peaks (around 93.77 and 93.85\,GHz) compared to two large outer DNP peaks (around 93.70 and 93.92\,GHz).
Owing to the broad DNP profiles, a slight variation between separate measurements exists (cf. 1.8\,K and 3.4\,K in Fig.~\ref{fig:Fig2}f as for 1.8\,K two measurements are shown and three for 3.4\,K).

\clearpage

\section{Longitudinal-detected (LOD) electron paramagnetic resonance (EPR)} \label{sec:SI_LOD}

\subsection{LOD profiles at 7\,T of 10\,\textmu m diamonds}

\begin{figure}[hbt]
	\centering
	\includegraphics[width=\linewidth]{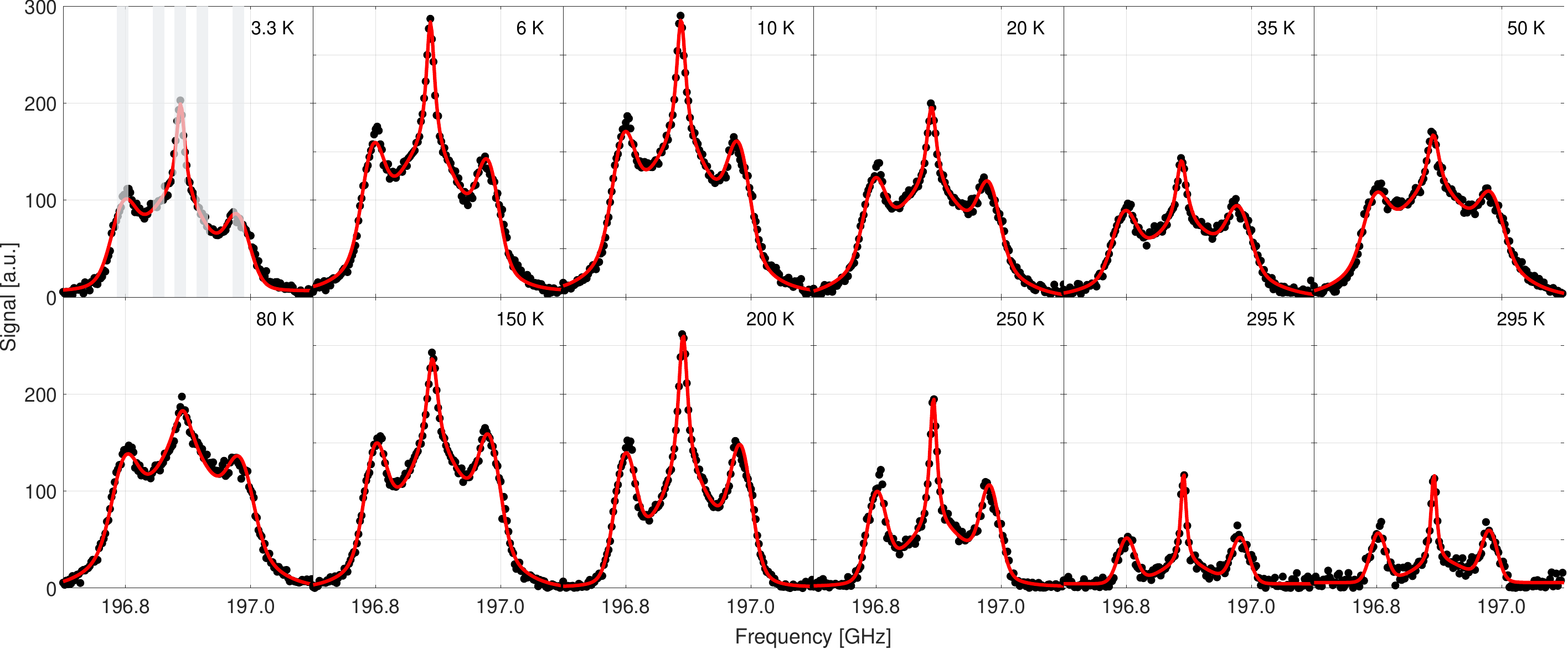}
	\caption{LOD profiles of the $10 \pm 2$\,\textmu m diamonds at different temperatures. 
    The profiles are fitted with a combination of P1 centers (Lorentzian central $m_I = 0$ line shape and Gaussian $m_I = \pm 1$ line shape) and a broad Gaussian line (cf. Eq.~\eqref{eq:EPRmodel} of the main text).
    The fit parameters for these are summarized in Fig.~\ref{fig:Fig3} of the main text.
    In each of the grey shaded frequency intervals, measured $\tau_\mathrm{LOD}$ are averaged and summarized in Fig.~\ref{fig:SI_electronTimes} (cf. text in Supplementary Material for more details).}
	\label{fig:SI_LODprofiles}
\end{figure}

The first type of LOD experiments presented in this study is the MW frequency dependence of the LOD signal (LOD profile or spectrum).
Fig.~\ref{fig:SI_LODprofiles} shows the measured LOD profiles between 3.4\,K and 295\,K fitted with Eq.~\eqref{eq:EPRmodel} of the main text consisting of a P1 centers and a broad component (for more details, see main text).
The fit parameters are summarized in Fig.~\ref{fig:Fig3} of the main part.

\begin{figure}[hbt]
	\centering
	\includegraphics[width=\linewidth]{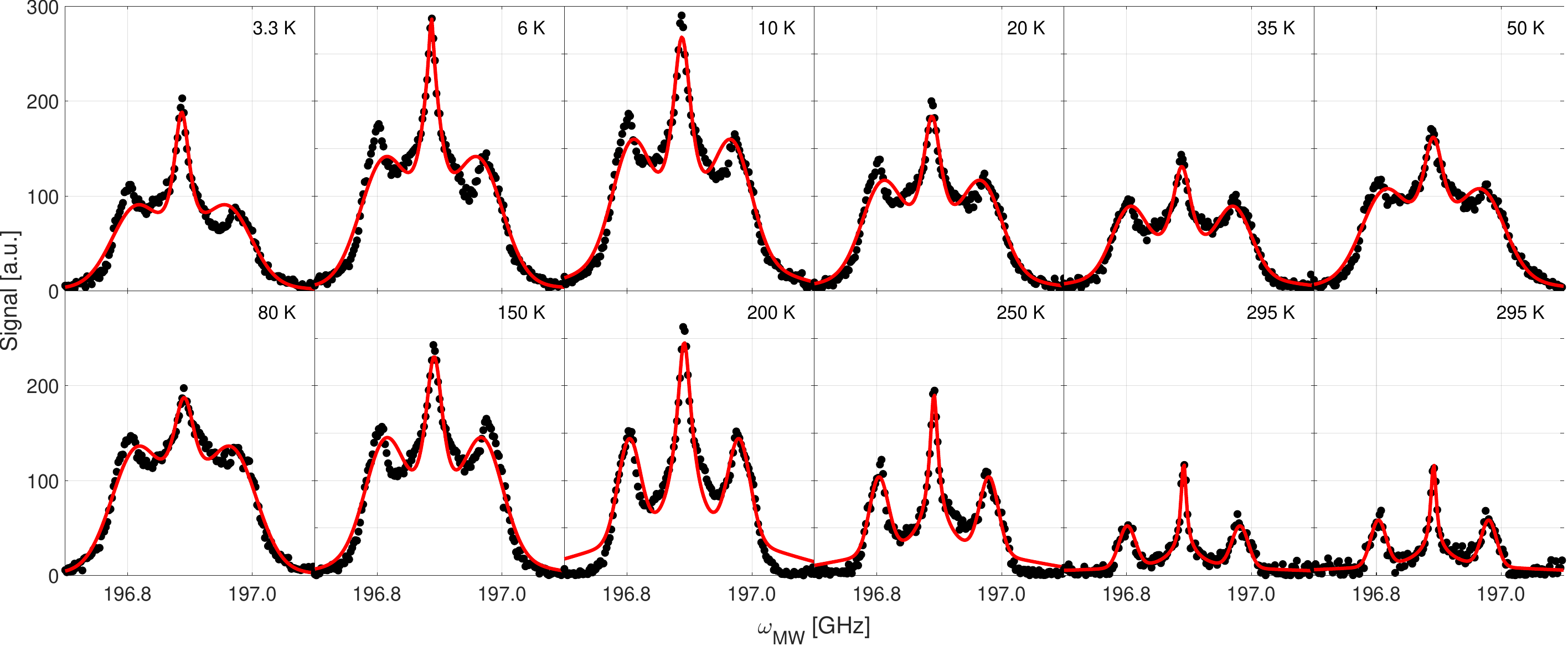}
	\caption{LOD profiles of the $10 \pm 2$\,\textmu m diamonds at different temperatures. 
    The profiles are fitted with a combination of two different types of P1 centers (Lorentzian central $m_I = 0$ line shape and Gaussian $m_I = \pm 1$ line shape) as described by Eq.~\eqref{eq:EPRmodel_expandedP1}: narrow isolated and cluster-broadened broad P1 centers as suggested in Ref.~\cite{bussandri_p1_2023,nir-arad_nitrogen_2023,palani_dynamic_2024} .
    The fit parameters for these are summarized in Fig.~\ref{fig:SI_LODprofilesExpandedFitFunctionFitParameters}.}
	\label{fig:SI_LODprofilesExpandedFitFunction}
\end{figure}

\begin{figure}[hbt]
	\centering
	\includegraphics[width=\linewidth]{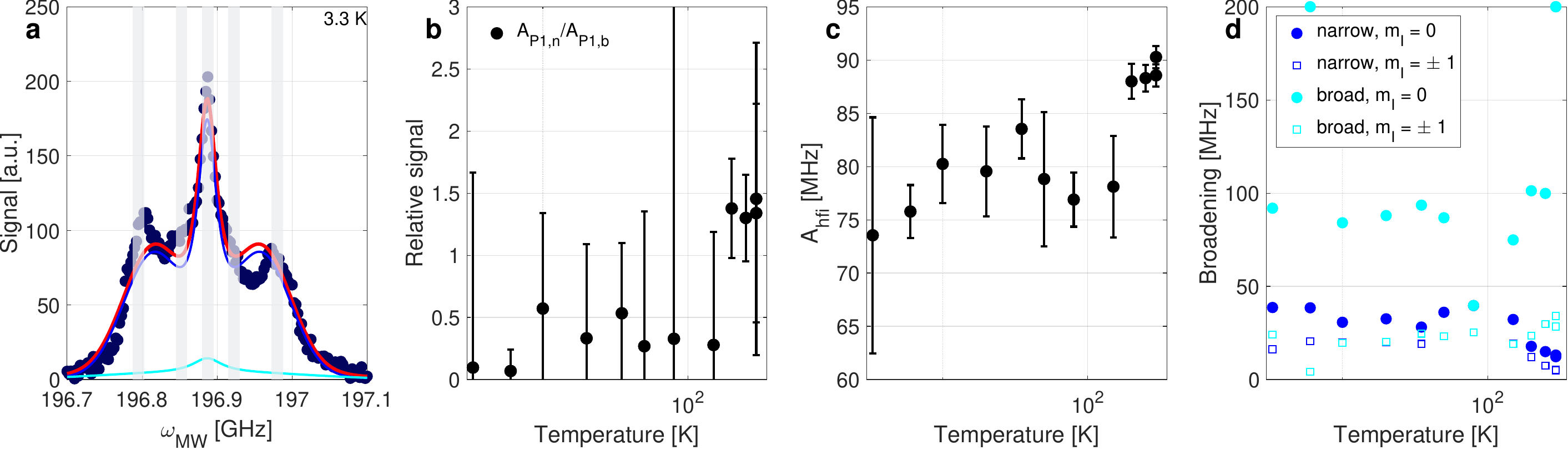}
	\caption{Summary of the fit parameters from Fig.~\ref{fig:SI_LODprofilesExpandedFitFunction} of the $10 \pm 2$\,\textmu m diamonds at different temperatures. 
    The profiles are fitted with a combination of two different types of P1 centers (Lorentzian central $m_I = 0$ line shape and Gaussian $m_I = \pm 1$ line shape) as described by Eq.~\eqref{eq:EPRmodel_expandedP1}: narrow isolated and cluster-broadened broad P1 centers as suggested in Ref.~\cite{bussandri_p1_2023,nir-arad_nitrogen_2023,palani_dynamic_2024}.}
	\label{fig:SI_LODprofilesExpandedFitFunctionFitParameters}
\end{figure}

In addition, the measured LOD EPR data was fitted with an approach motivated by isolated and clustered P1 centers as discussed in \cite{bussandri_p1_2023,nir-arad_nitrogen_2023,palani_dynamic_2024} leading to a narrow and broad P1 spectrum as discussed in the main text, particularly Eq.~\eqref{eq:EPRmodel_expandedP1}.
The fitted spectra in the range of 3.3\,K to 295\,K are shown in Fig.~\ref{fig:SI_LODprofilesExpandedFitFunction} with the fit parameters summarized in Fig.~\ref{fig:SI_LODprofilesExpandedFitFunctionFitParameters}.

\FloatBarrier

\subsection{LOD decay times at 7\,T of 10\,\textmu m diamonds}

\begin{figure}[hbt]
	\centering
	\includegraphics[width=\linewidth]{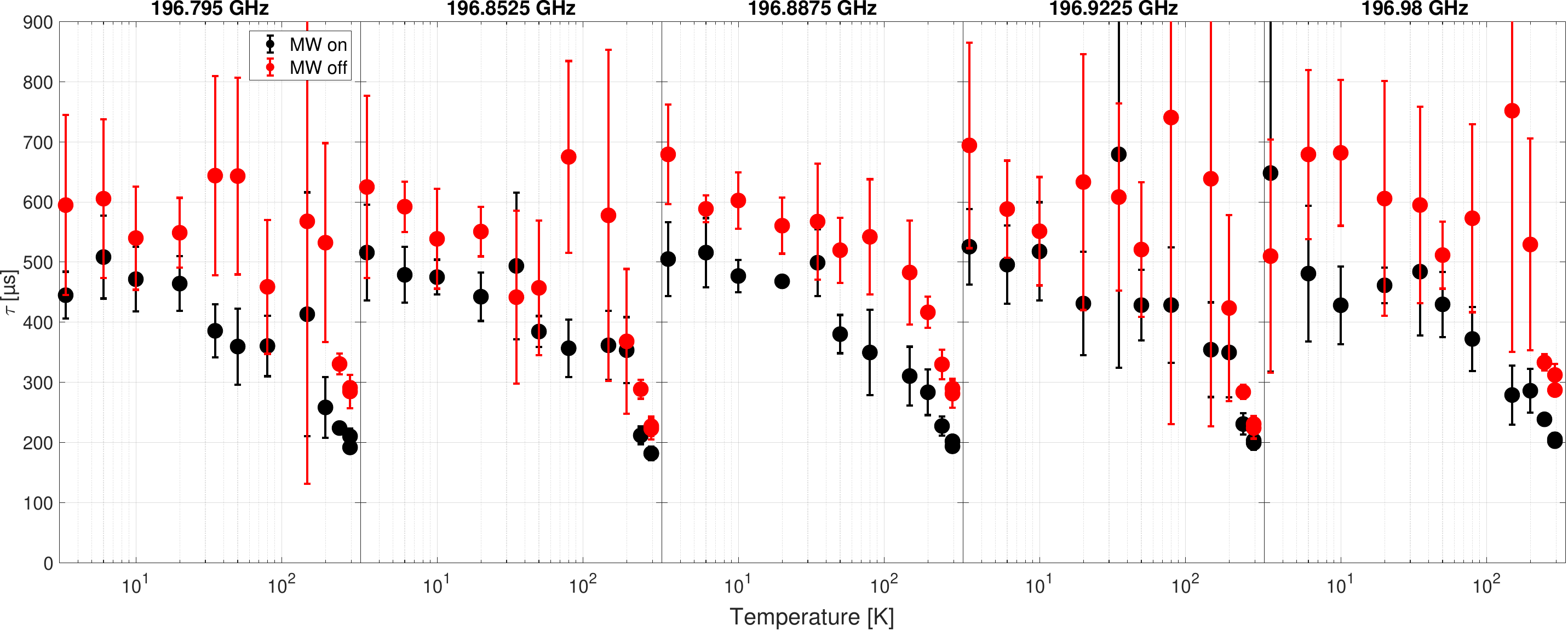}
	\caption{Electronic relaxation times $\tau_\mathrm{LOD}$ in LOD EPR at 7\,T for different temperatures and frequency intervals sketched by the grey shading in Fig.~\ref{fig:Fig3}a. 
    Each data point is an average over five frequencies. Uncertainties are derived from the standard deviation of the five frequencies.}
	\label{fig:SI_electronTimes}
\end{figure}

LOD EPR measures the induced voltage after the polarization of the electrons is changed either upon switching the MW on or off (this has similarities to a $T_{2,\mathrm{n}}^*$ experiment in NMR).
Our LOD EPR experiments consist of two parts: (i) The MW is switched on, saturating the electron line at the given frequency and through electron spectral diffusion (eSD) to neighbouring frequencies. 
(ii) The MW is switch off and the electron system relaxes back to thermal equilibrium. 
These two parts are denoted as 'MW on' and 'MW off' in Fig.~\ref{fig:SI_electronTimes}.
The signal decays in $<1$\,ms (cf. Fig.~\ref{fig:SI_electronTimes} ) - consistent with other materials used in DNP \cite{himmler_electroplated_2022}. 
The MW on and off parts are each fitted with a mono-exponential decay with a decay constant $\tau_\mathrm{LOD}$. 
Owing to the rather low SNR, the values of $\tau_\mathrm{LOD}$ for a single data point fluctuate.
Averaging over a narrow frequency interval gives more reliable values.

We note that LOD EPR does not measure $T_{1,\mathrm{e}}$ but rather $\tau_\mathrm{LOD} = \left(\frac{1}{T_{1,\mathrm{e}}}+\frac{1}{\tau_\mathrm{ff}}\right)^{-1}$ with $\tau_\mathrm{ff}$ being an electronic flip-flop time describing the electron couplings and with it electron spectral diffusion.

For five different frequency intervals which are shaded in gray in Fig.~\ref{fig:SI_LODprofiles} for 3.3\,K, the obtained averaged LOD time constants $\tau_\mathrm{LOD}$ are shown in Fig.~\ref{fig:SI_electronTimes}.
The trend is for all five frequencies similar: At low temperatures $\tau_\mathrm{LOD}$ is around 500-600\,\textmu s with the 'MW on' time slightly shorter. 
$\tau_\mathrm{LOD}$ is rather stable up to temperatures of a few tens of Kelvin before shortening to around 200\,\textmu s at room temperature.
The measured LOD time constants appear frequency independent (cf. Fig.~\ref{fig:SI_electronTimes}), indicating electron-electron cross relaxation between the P1 and broad electron lines.
Decreasing the temperature from room temperature to liquid-helium temperatures monotonically increases $\tau_\mathrm{LOD}$ from around 200-300\,\textmu s to around 500-600\,\textmu s.
Assuming that $T_{1,\mathrm{e}} \gg \tau_\mathrm{LOD}$ at liquid-helium temperatures, the measured time constant would describe the electronic flip-flop time $\tau_\mathrm{ff}$.
For $\tau_\mathrm{LOD} \approx 300$\,\textmu s at room temperature and $\tau_\mathrm{ff} \approx 600$\,\textmu s from the low temperature LOD EPR data (cf. Fig.~\ref{fig:SI_electronTimes}), $T_{1,\mathrm{e}} \approx 600$\,\textmu s at room temperature which is shorter than the typically reported $\approx2$\,ms \cite{terblanche_room-temperature_2000,terblanche_13c_2001} for P1 centers at room temperature.

\FloatBarrier

\subsection{Power dependence at 7 T of 10\,\textmu m diamonds}

\begin{figure}[ht]
	\centering
	\includegraphics[width=\linewidth]{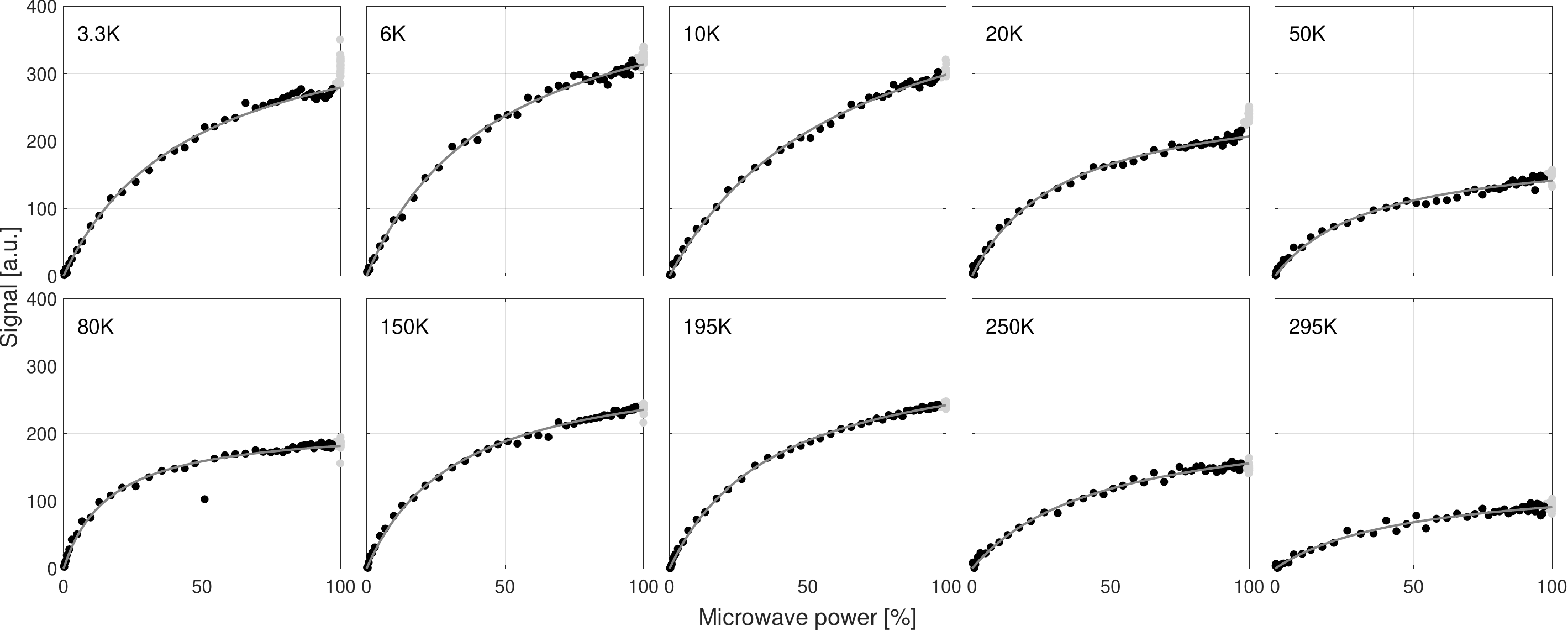}
	\caption{Power dependence of the LOD EPR signal at 7\,T, 196.905\,GHz for different temperatures. 
    The data is fitted with $a\left(1-\frac{1}{bx+1}\right)$ (cf. Eq.~\eqref{eq:TorreySimplified}) with $bx = \gamma_e^2 B_{1,\mathrm{MW}}^2 T_{2,\mathrm{e}} T_{1,\mathrm{e}}$  being the saturation factor ($x$ is the MW power which defines $B_{1,\mathrm{MW}}$)  \cite{von_witte_two-electron_2024}. 
    The best fit parameters are shown in Fig.~\ref{fig:SI_TorreyParameters}.}
	\label{fig:SI_TorreyPower}
\end{figure}

Another group of LOD EPR measurements concerns the power dependence of the signal.
LOD EPR power measurements for temperatures between 3.3\,K and 295\,K are shown in Fig.~\ref{fig:SI_TorreyPower}.
For the power curves, the measured signal of 'MW off' part was numerically integrated (summed over the data points).
The resulting power curves are fitted with the time-independent part of the Torrey model of damped Rabi oscillations which is equivalent to the z-component of the time-independent Bloch equations \cite{von_witte_two-electron_2024}.
Specifically, the model is given by
\begin{align}
    1-\frac{P_{e,\infty}}{P_{0,e}} = 1- \frac{1}{\gamma_e^2B_{1,\mathrm{MW}}^2T_{2,\mathrm{e}}T_{1,\mathrm{e}}+1}  \label{eq:TorreySimplified}
\end{align}
with the electronic relaxation times $T_{1,\mathrm{e}}$ (spin-lattice) and $T_{2,\mathrm{e}}$ (spin-spin).
$B_{1,\mathrm{MW}}$ is the magnetic field generated by the MW perpendicular to the main magnetic field $B_0$.

\begin{figure}[hbt]
	\centering
	\includegraphics[width=\linewidth]{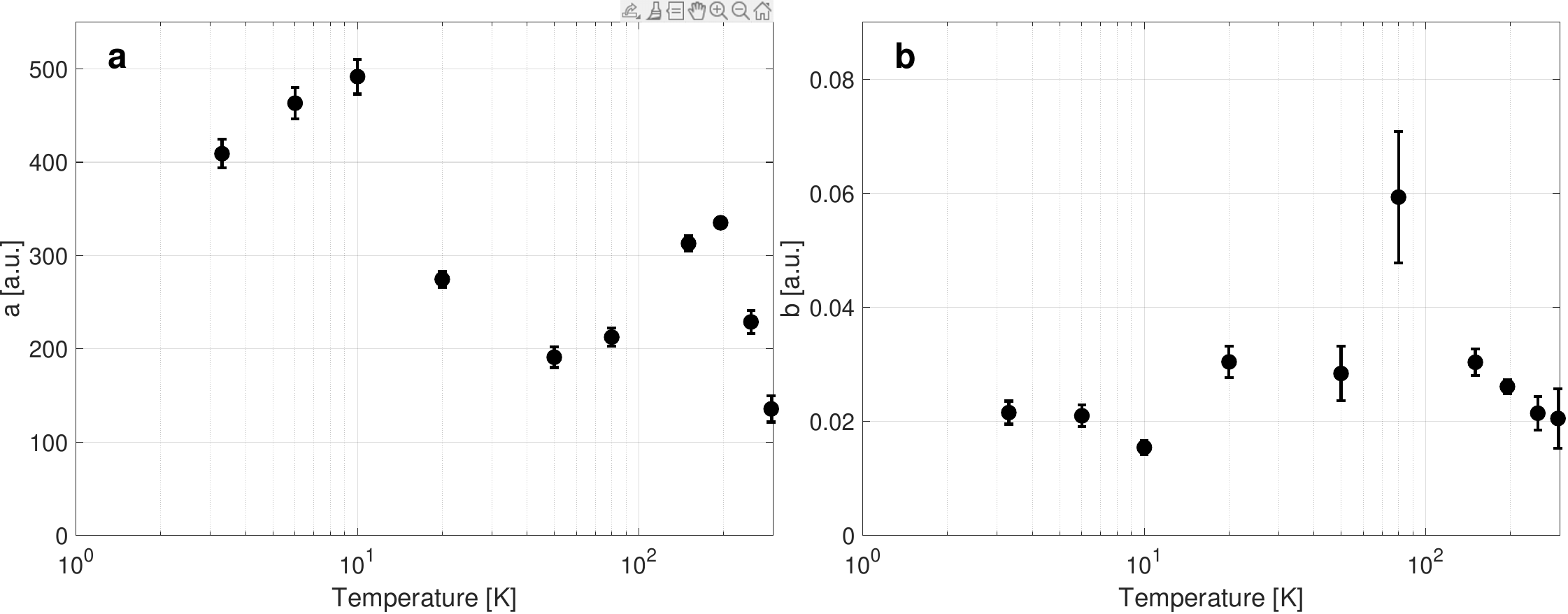}
	\caption{Best fit parameters of the power dependence of the LOD EPR signal at 7\,T, 196.905\,GHz for different temperatures as shown in Fig.~\ref{fig:SI_TorreyPower}. 
    The data is fitted with $a\left(1-\frac{1}{bx+1}\right)$, derived from the simplified Torrey model (cf. Eq.~\eqref{eq:TorreySimplified}) \cite{von_witte_two-electron_2024}. 
    The data points for the highest MW powers are omitted (shown in light gray) for the fitting process as problems with setting the MW occurred at the beginning of each measurement (the highest powers were measured first).
    Uncertainties might be smaller than the symbols.}
	\label{fig:SI_TorreyParameters}
\end{figure}

If the conversion from MW power to $B_{1,\mathrm{MW}}$ for a given experimental set-up is known, LOD EPR might be used to measure $T_{1,\mathrm{e}}$ and $T_{2,\mathrm{e}}$ as relevant for DNP.
In the current case, a simplified expression was used (cf. captions of Figs.~\ref{fig:SI_TorreyPower} and \ref{fig:SI_TorreyParameters}) and the best fit parameters are shown in Fig.~\ref{fig:SI_TorreyParameters}.
We note that the Torrey model is able to explain the nearly linear LOD EPR signal-power data found in Ref.~\cite{himmler_electroplated_2022}.

The amplitude of the signal does not follow the thermal electron polarization, which increases by around two orders of magnitude between the highest and lowest temperature.
In contrast, the amplitude of the LOD signal only increases around fourfold. 
At high temperatures, the applied MW power is almost enough to fully saturate the electron line (the measured signal is close to the fitted maximum signal) while at low temperatures eventually only around half the electron line is saturated (at 10\,K the saturation seems the most difficult).
The $b$ parameter describing the electron relaxation is nearly independent of the temperature with values around 0.02-0.03 for all temperatures except 80\,K.

\FloatBarrier

\subsection{LOD profiles of nanodiamonds}

\begin{figure}[ht]
	\centering
	\includegraphics[width=0.7\linewidth]{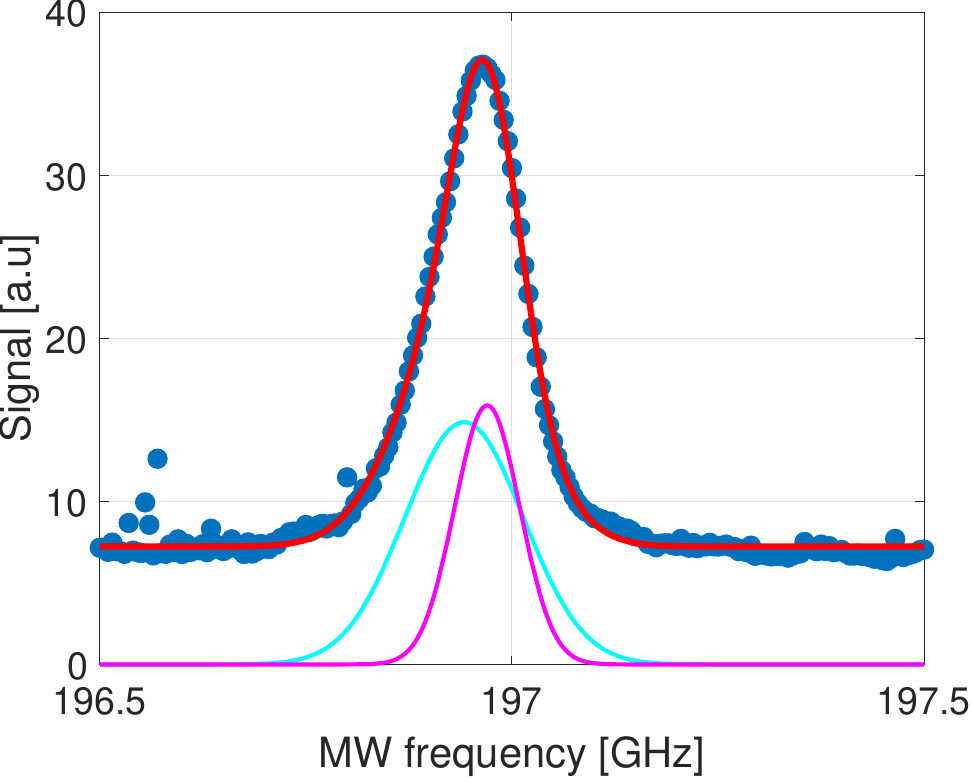}
	\caption{LOD EPR spectrum of nanodiamonds ($<10$\,nm) sample at 7\,T and 20\,K.
    The spectrum is fitted with Eq.~\eqref{eq:EPRmodel_nanodiamonds}.
    The fit parameters can be found in Tab.~\ref{tab:SI_LOD_7T_nanodiamonds}. }
	\label{fig:SI_LOD_7T_nanodiamonds}
\end{figure}

Fig.~\ref{fig:SI_LOD_7T_nanodiamonds} shows the fitted LOD profile of a nanodiamond sample ($<10$\,nm) at 7\,T and 20\,K. 
We immediately note the absence of the three P1 peaks as expected for small nanodiamonds with with a large surface-to-volume ratio.
To fit the measured spectrum, we use
\begin{align}
    S_\mathrm{EPR} &= \frac{S_\mathrm{b}}{\sqrt{2\pi}\sigma_\mathrm{b}} e^{\frac{\left(\nu-\nu_\mathrm{b}\right)^2}{2\sigma_\mathrm{b}^2}} + \frac{S_\mathrm{n}}{\sqrt{2\pi}\sigma_\mathrm{n}} e^{\frac{\left(\nu-\nu_\mathrm{n}\right)^2}{2\sigma_\mathrm{n}^2}} + S_\mathrm{offset} \label{eq:EPRmodel_nanodiamonds} 
\end{align}
with a broad and narrow spin-1/2 Gaussian electron line.
Details about the fit parameters are found in Tab.~\ref{tab:SI_LOD_7T_nanodiamonds}.
The broad and narrow spin-1/2 components are discussed in more detail below.

\begin{table}[hbt]
\begin{center}
\begin{tabular}{l|l|l}
    & broad & narrow \\ \hline
    Weight [\%] & 63(12) & 37(12) \\
    Frequency [GHz] & 196.942(6) & 196.970(3) \\ 
    Line Width [MHz] & 72(5) & 40(4) \\
\end{tabular}
\end{center}
\caption{Fit parameters of the LOD spectrum shown in Fig.~\ref{fig:SI_LOD_7T_nanodiamonds} and fitted with Eq.~\eqref{eq:EPRmodel_nanodiamonds}.}
\label{tab:SI_LOD_7T_nanodiamonds}
\end{table}

\FloatBarrier

\begin{figure}[ht]
	\centering
	\includegraphics[width=0.7\linewidth]{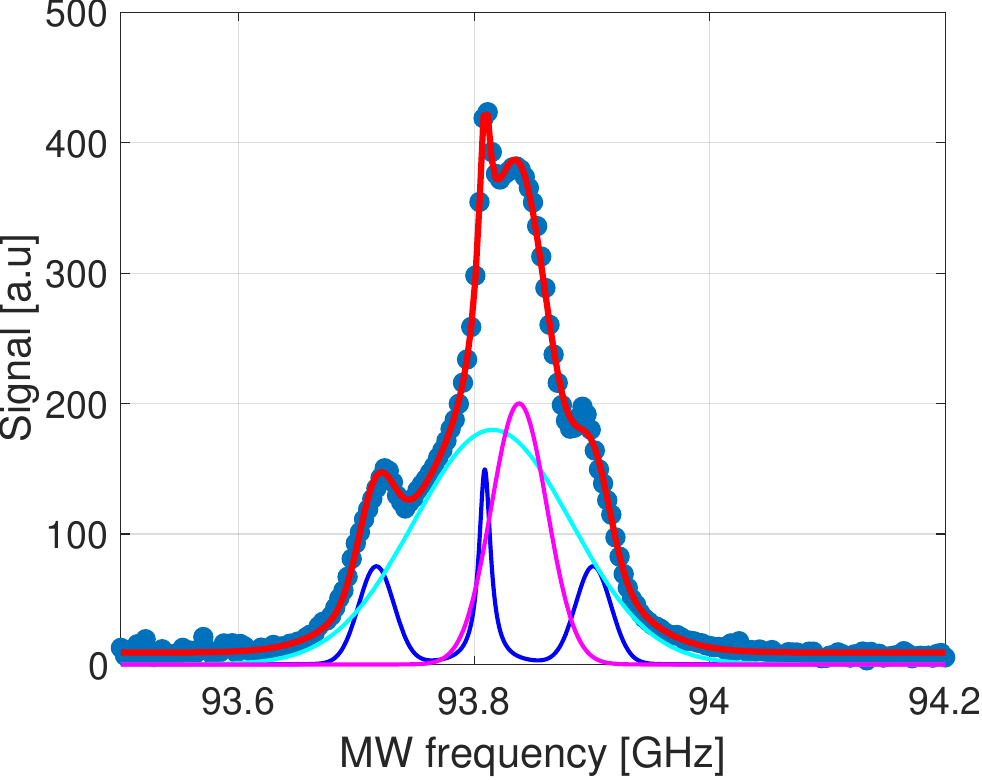}
	\caption{LOD EPR spectrum of diamonds smaller than 250\,nm studied at 3.4\,T and around 20\,K.
    The spectrum is fitted with Eq.~\eqref{eq:EPRmodel_extraLine}.
    The fit parameters can be found in Tab.~\ref{tab:SI_LOD_3T_250nm}. }
	\label{fig:SI_LOD_3T_250nm}
\end{figure}

Fig.~\ref{fig:SI_LOD_3T_250nm} shows the fitted LOD profile of diamonds smaller than 250\,nm at 3.4\,T and 20\,K. 
In contrast to the larger 10\,\textmu m diamonds at 7\,T, an additional rather narrow electron line emerges at a frequency roughly 25\,MHz higher than the $m_I = 0$ line of the P1 center.
To fit the measured spectrum, we added another term to Eq.~\eqref{eq:EPRmodel} to describe the rather narrow line (denoted with "n" subscript)
\begin{align}
    S_\mathrm{EPR} &= \frac{S_\mathrm{P1}}{\sqrt{2\pi}\sigma_{\pm 1}} \left[e^{-\frac{\left(\nu - (\nu_\mathrm{P1} - A_\mathrm{P1})\right)^2}{2\sigma_{ \pm 1}^2}} 
    + e^{-\frac{\left(\nu - (\nu_\mathrm{P1} + A_\mathrm{P1})\right)^2}{2\sigma_{ \pm 1}^2}} \right]
    + \frac{S_\mathrm{P1}}{\pi}\frac{\sigma_0}{\left(\nu - \nu_\mathrm{P1}\right)^2 + \sigma_0^2} 
    + \frac{S_\mathrm{b}}{\sqrt{2\pi}\sigma_\mathrm{b}} e^{\frac{\left(\nu-\nu_\mathrm{b}\right)^2}{2\sigma_\mathrm{b}^2}} + 
    + \frac{S_\mathrm{n}}{\sqrt{2\pi}\sigma_\mathrm{n}} e^{\frac{\left(\nu-\nu_\mathrm{n}\right)^2}{2\sigma_\mathrm{n}^2}} +S_\mathrm{offset} \label{eq:EPRmodel_extraLine} 
\end{align}
Details about the fit parameters are found in Tab.~\ref{tab:SI_LOD_3T_250nm}.
Similar to 7\,T, the broad component is the largest contributor to the signal at low temperatures, is centered within a few MHz of the $m_I = 0$ P1 line and has a similar line width at 3.4\,T. 
The results for the broad and narrow spin-1/2 component in terms of center frequencies and line widths are similar to $<10$\,nm diamonds at 7\,T (cf.~Fig.~\ref{fig:SI_LOD_7T_nanodiamonds} and Tab.~\ref{tab:SI_LOD_3T_250nm}).
However, the $<10$\,nm diamonds with a much higher surface-to-volume ratio have a more similar intensity ratio between the broad and narrow components.
Furthermore, the narrow component is not observed for the larger 10\,\textmu m diamonds.
This suggests that the narrow spin-1/2 component arises from surface defects while the broad spin-1/2 component arises from a different type of bulk defect than P1 centers.

\begin{table}[hbt]
\begin{center}
\begin{tabular}{l|c|c|c}
    & P1 & broad & narrow \\ 
    $A_\mathrm{P1}$ [MHz] & 92.1(7) &  &  \\ \hline
    Weight [\%] &  6.4(7) & 68(3) & 25(3) \\ 
    Frequency [GHz] & 93.809(3) & 93.816(2) & 93.838(1) \\ 
    Line Width [MHz] & 15.0(9) ($m_I = \pm 1$), 6.0(5) ($m_I = \pm 0$) & 67(2) & 24(1) \\
\end{tabular}
\end{center}
\caption{Fit parameters of the LOD spectrum shown in Fig.~\ref{fig:SI_LOD_3T_250nm} and fitted with Eq.~\eqref{eq:EPRmodel_extraLine}.}
\label{tab:SI_LOD_3T_250nm}
\end{table}

\clearpage

\section{X-band EPR} \label{sec:SI_X-band}

\begin{figure}[ht]
	\centering
	\includegraphics[width=\linewidth]{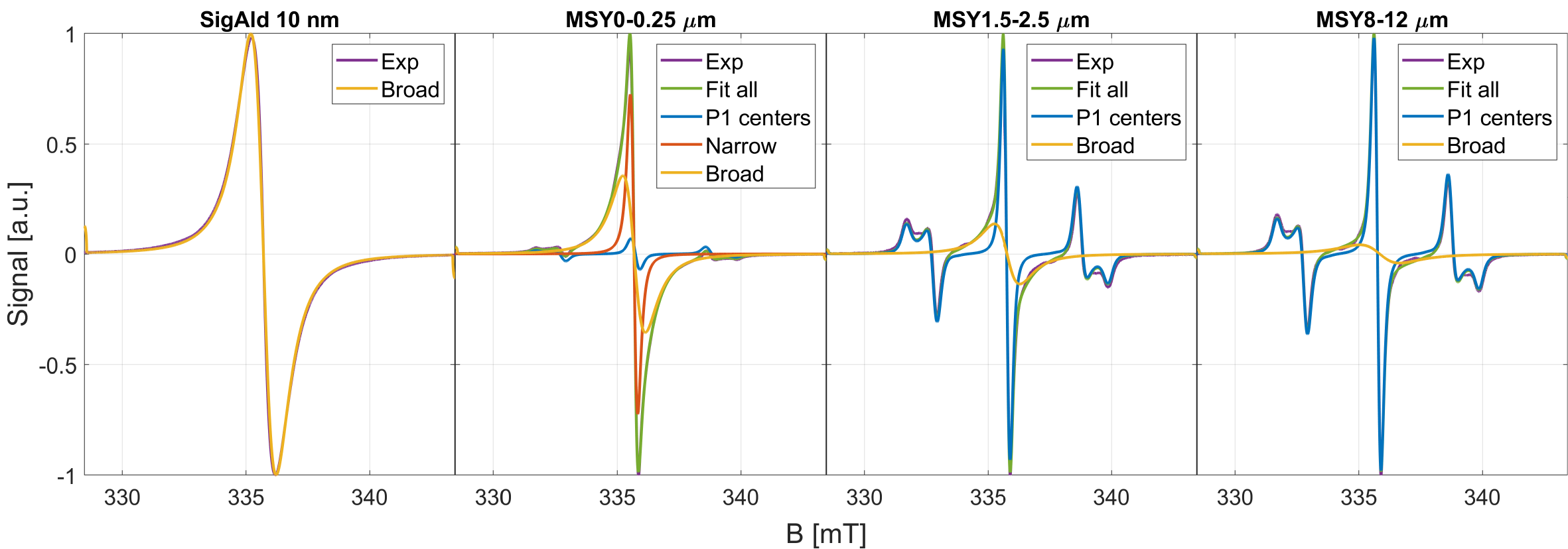}
	\caption{Measured and fitted X-band EPR spectra of four different diamond samples at room temperature. 
    The spectra are fitted with EasySpin and a combination of broad and narrow spin-1/2 defects as well as P1 centers.
    The assumed spin system for the fit models is described by the legends.}
	\label{fig:SI_EPR_Konstantin_fits}
\end{figure}

\begin{figure}[ht]
	\centering
	\includegraphics[width=\linewidth]{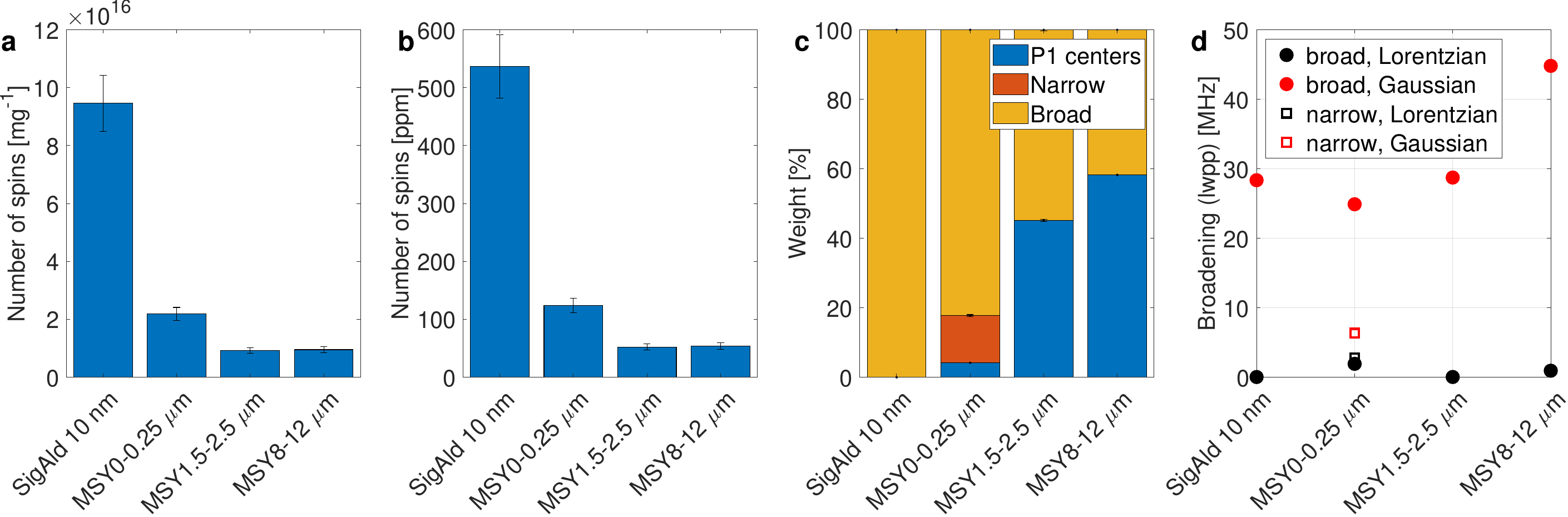}
	\caption{\textbf{(a)} Number of electron spins per mg or \textbf{(b)} converted to ppm in the different diamond samples.
    \textbf{(c)} Relative weights of the different spin systems in percent of the total number of spins as displayed in a,b.
    The assumed spin system for the fit models is described by the legends in Fig.~\ref{fig:SI_EPR_Konstantin_fits}.
    \textbf{(d)} Lorentzian and Gaussian line broadening of the narrow and broad components.}
	\label{fig:SI_EPR_Konstantin_NumberOfSpins_Ratios}
\end{figure}

In addition to the LOD EPR data, we performed X-band EPR (335\,mT) to estimate the number of electron spins and to enable a comparison to previous works on diamond particles \cite{rej_hyperpolarized_2015, waddington_phase-encoded_2019,boele_tailored_2020}.
For this, we compared the $10\pm2$\,\textmu m sample used throughout this work with $2\pm0.5$\,\textmu m, $<250$\,nm and $<10$\,nm diamond particles (cf. Sec.~\ref{sec:SI_LOD}).
The measured X-band EPR spectra are shown in Fig.~\ref{fig:SI_EPR_Konstantin_fits}.
The spectra are fitted with EasySpin \cite{stoll_easyspin_2006} with a combination of broad and narrow spin-1/2 defects as well as P1 centers similar to Ref.~\cite{rej_hyperpolarized_2015, waddington_phase-encoded_2019,boele_tailored_2020}.
The total number of electron spins with respect to TEMPO and porous silicon samples are shown in Fig.~\ref{fig:SI_EPR_Konstantin_NumberOfSpins_Ratios}.
The spin concentration decreases with increasing particle size.
With increasing particle sizes, the fraction of P1 centers increases.
For the 10\,\textmu m diamonds, an estimated 58\% of 54\,ppm of electron spins are P1 centers.
We notice that the Gaussian broadening in the $10 \pm 2$\,\textmu m-sized diamonds is noticeably larger than in the other samples (cf. Fig.~\ref{fig:SI_EPR_Konstantin_NumberOfSpins_Ratios}d) and appears in agreement with the broadening of the broad component in LOD EPR at 295\,K (cf. Fig.~\ref{fig:Fig3}f).

\FloatBarrier

\clearpage

\section{Additional information about possible defects} \label{sec:SI_defectDiscussion}

In the following, we will discuss a selected number of nitrogen-based bulk defects in diamond that could be consistent with the results presented in this work.
Specifically, we will discuss three possible nitrogen defects previously studied together with the P1 center which posses similar $g$-factors as the P1 or at least overlapping EPR lines at X- ($\approx0.3$\,T) or W-band ($\approx 3.4$\,T) as possible candidates to explain the broad (spin-1/2) component.
We note that a much larger range of defects in diamond is known and Ref.~\cite{loubser_electron_1978} might provide a first overview to the interested reader although information on some of the defects got refined in subsequent years.
Below, we follow the nomenclature from Refs.~\cite{reynhardt_temperature_1998,terblanche_13c_2001,reynhardt_spin_2003} in naming the different defects as some ambiguity in naming between different publications exists.

\begin{itemize}
    \item The first nitrogen defect we discuss in some more detail besides the P1 center is the so-called P2 center which consists of three substituional nitrogen atoms and a vacancy (N\textsubscript{3}V).
    The P2 center has a broad EPR line owing to its large number of energy levels and overlaps only with the P1's $m_I = 0$ EPR line at X- and W-band \cite{terblanche_room-temperature_2000, reynhardt_spin_2003}.
    The P2 center has a long and field-independent $T_{1,\mathrm{e}}^\mathrm{P2} \approx 2.2$\,ms at room temperature \cite{terblanche_room-temperature_2000} which is similar to the $T_{1,\mathrm{e}}^\mathrm{P1}$ at room temperature \cite{reynhardt_temperature_1998}.
    If P1 and P2 centers are present in the same sample, they tend to cross-relax such that both have a similar temperature dependence with $T_{1,\mathrm{e}}$ on the order of a few milliseconds at room temperature and seconds to hundreds of seconds below 10\,K \cite{reynhardt_temperature_1998}.
    At low temperatures, a higher P1 concentration, i.e. 0.2 vs. 95\,ppm, leads to a shorter $T_{1,\mathrm{e}}$ on the order of several seconds \cite{reynhardt_temperature_1998}.

    \item The N2 center (called W7 center in Ref.~\cite{loubser_electron_1978}) is considered to consist of a N-C-N or N-C-C-N complex \cite{loubser_singly_1973,loubser_electron_1978,terblanche_13c_2001}.
    Its properties are temperature-dependent due to a dynamic Jahn-Teller effect with a rather low activation energy \cite{loubser_singly_1973} leading to an estimated room temperature spin-lattice relaxation time of a few nanoseconds \cite{terblanche_13c_2001}.
    The N2 center shows a broad central line and six weaker hyperfine lines on either side of the central line \cite{loubser_singly_1973,loubser_electron_1978}.

    \item The N3 center consists of substitutional N and O atoms \cite{wyk_endor_1992}) with a slight $g$-factor anisotropy at room temperature \cite{wyk_endor_1992} which could explain the rather broad electron line observed in our LOD experiments.
    Strong cross-relaxation with P1 centers was previously observed \cite{reynhardt_temperature_1998} causing an unexplained shortening of $T_{1,\mathrm{e}}^\mathrm{P1}$ with a pronounced asymmetry between different P1 hyperfine contributions reaching $T_{1,\mathrm{e}}^{\mathrm{P1},~m_I=0}/T_{1,\mathrm{e}}^{\mathrm{P1},~m_I=-1} \approx 100$ below 10\,K \cite{reynhardt_temperature_1998}.   
\end{itemize}

Measurements at 4.7\,T (200\,MHz \textsuperscript{1}H Larmor frequency, 50\,MHz \textsuperscript{13C} Larmor frequency) and room temperature \cite{terblanche_13c_2001,reynhardt_spin_2003} suggest that a combination  P1 and P2 centers is inefficient in relaxing nuclear polarization even at concentrations of around 5\,ppm ($T_{1,\mathrm{n}} > 10$\,h) while a mixture of P1 centers and N2 (N3) centers with 0.04 (10)\,ppm leads to $T_{1,\mathrm{n}}$ around 5.4 (1.4)\,h. 

Based on these considerations, the broad line detected upon cooling in our LOD EPR experiments could be due to N2 or N3 centers although other defects cannot be completely ruled out.
Both defects might explain a broad line around the $m_I = 0$ P1 EPR line, a fast relaxation at room temperature and possibly at low temperatures (sub-second time scale).

\clearpage

\section{Uncoupled compartments} \label{sec:SI_uncoupledCompartments}


We start with a recap of the previously introduced single homogeneous compartment model \cite{von_witte_modelling_2023}:
The hyperpolarization build-up can be described through a first-order differential equation with a hyperpolarization injection rate constant $k_\mathrm{W}$ and a relaxation rate constant of the build-up $k_\mathrm{R}^\mathrm{bup}$ 
\begin{align}
	\frac{\text{d}P}{\text{d}t} = (A-P) k_\mathrm{W} - k_\mathrm{R}^\mathrm{bup} P \label{eq:ODE_bup_1compartment}
\end{align}
with $A$ describing the theoretical maximum of hyperpolarization achievable, i.e., the thermal electron polarization in DNP.
The solution of Eq.~\eqref{eq:ODE_bup_1compartment} is a mono-exponential curve which can be compared with the phenomenological description of the build-up curve by $P(t) = P_0 (1-e^{-t/\tau_\mathrm{bup}})$ to express the experimental parameters in terms of model parameters. 
Here, $P_0$ is the steady-state polarization and $\tau_{\mathrm{bup}}$ the build-up time.
\begin{subequations} \label{eq:1compartment_bup_solution}
	\begin{align}
		\tau_{\mathrm{bup}}^{-1} &= k_\mathrm{W}+k_\mathrm{R}^\mathrm{bup} \label{eq:1compartment_bup_tau} \\
		P_0 &= \frac{Ak_\mathrm{W}}{k_{W}+k_\mathrm{R}^\mathrm{bup}} = Ak_\mathrm{W}\tau_{\mathrm{bup}} \label{eq:1compartment_bup_P0}
	\end{align}
\end{subequations}
 
For the decay, $k_\mathrm{W}$ would be set to zero (MW off), leading to $\tau_\mathrm{decay}^{-1}=k_\mathrm{R}^\mathrm{decay}$. 

Extending the one-compartment model to two uncoupled compartments with separate injection and relaxation rates is straightforward. 
Such a situation might be realized for a material consisting of two phases with different compositions (radical concentration, NMR-active spin concentration) such that each compartment follows its own mono-exponential build-up (cf. Eqs.~\eqref{eq:ODE_bup_1compartment}, \eqref{eq:1compartment_bup_solution}).
Crucially, spin diffusion between the two compartments needs to be suppressed, e.g., through a resonance frequency difference rendering inter-compartment nuclear flip-flops energy non-conserving.
If the frequency difference between the two compartments is small compared to the NMR linewidth such that the two compartments cannot be clearly discriminated through different peaks, the total measured signal describes the total magnetization created in the two compartments.
In such a case, the resulting build-up would take a bi-exponential form

\begin{align}
	P &= P_{0,2}  \left[\alpha  \left(1-e^{-t/\tau_1}\right) + (1-\alpha)  \left(1-e^{-t/\tau_2}\right)\right] \nonumber \\
	&= P_{0,2}  \left[1 - \alpha  e^{-t/\tau_1} - (1-\alpha)  e^{-t/\tau_2}\right] \label{eq:DoubleExpBup_experiment}
\end{align}
with the relative weight of the two time constants $\alpha$.
Experimentally, four parameters are extracted from the build-up while in the theoretical model five parameters are required: two injection and relaxation rates each giving rise to the two steady-state polarizations and build-up times as well as the relative size of the compartments.
Hence, for two uncoupled compartments, it is difficult to extract information about the individual compartments based on the above compartment model ansatz.

\begin{figure*}[ht]
	\centering
	\includegraphics[width=\textwidth]{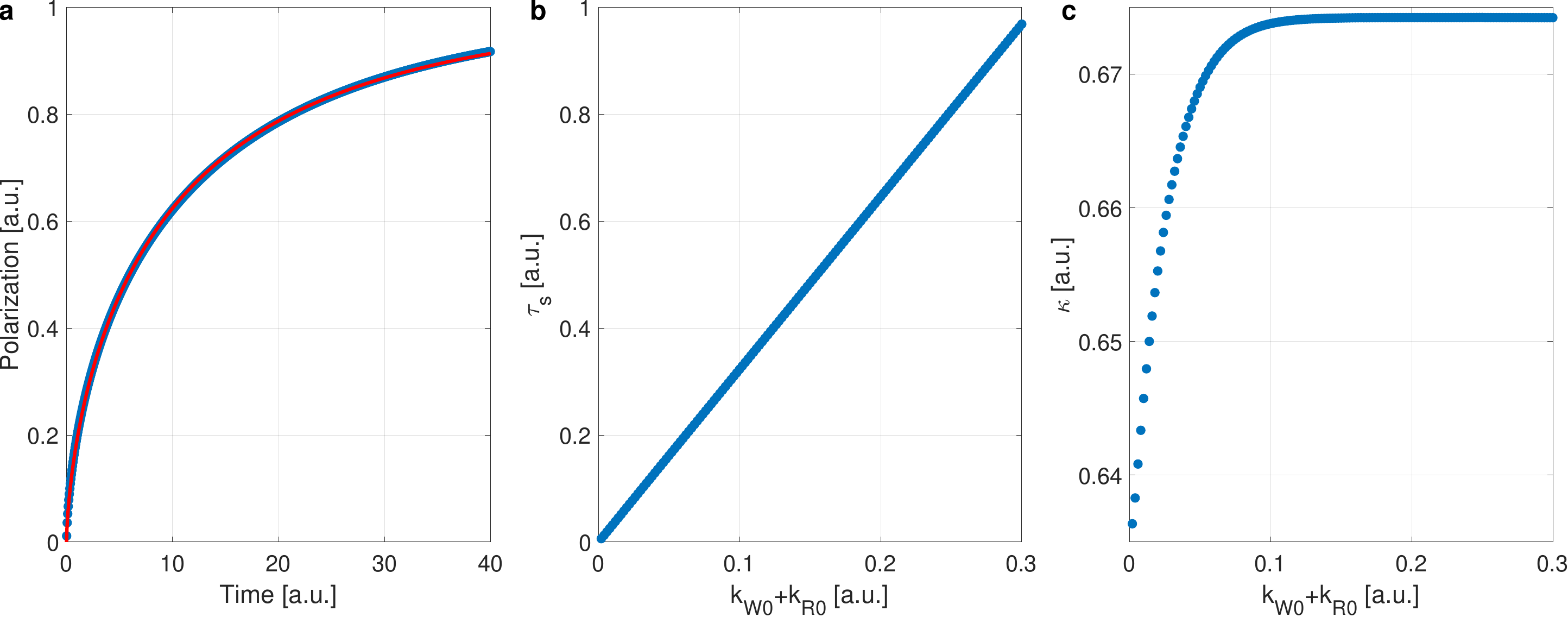}
	\caption{Numerical evaluation of Eq.~\eqref{eq:IncompleteGamma_Solution}  ($r_c=1$ and without the $P_{0,s}=\frac{Ak_{W0}}{k_{R0}+k_{W0}} \frac{4}{3}\pi \frac{1}{4}$ prefactor) and fitting the resulting curve with a stretched exponential. \textbf{a} A typical fit build-up and its stretched exponential build-up fit. Varying $k_{R0}+k_{W0}$ changes the characteristic time constant $\tau_S$ (\textbf{b}) and stretch exponent $\kappa$ of a stretched exponential as defined in equation \eqref{eq:StretchExp_SI}.
    For the linear fit of $\tau_\mathrm{s}$ a slope of $3.229\pm0.001$ is found.}
	\label{fig:Simulation_stretchExp}
\end{figure*}

However, for infinitely many uncoupled compartments with some additional assumptions, we can describe stretched exponential build-ups: 
It was recently proposed \cite{jardon-alvarez_enabling_2020} that for systems without spin diffusion and nuclear relaxation only through paramagnetic relaxation, the build-up can be described by a stretched exponential. 
In such a case the DNP transfer rate per lattice site and the relaxation scale with $r^{-6}$, with $r$ being the distance between the nuclear spin and the electron as both processes are mediated by the hyperfine coupling between the paramagnetic center and the nucleus under consideration. 
This case can be considered as infinitely many uncoupled compartment (with only paramagnetic relaxation).
Thus, our DNP injection rate has the same spatial scaling as the DNP transfer rate and the paramagnetic relaxation. We write for this case $k_{W1} = \frac{k_{W0}}{r^6}$ and $k_\mathrm{R}= \frac{k_{R0}}{r^6}$, ignoring any angular dependence of the hyperfine interaction. 
To describe the polarization build-up of the total system, we can divide it into systems with the same distance to the paramagnetic center, treat these as single compartments and average over all subsystems. 
For this we find
\begin{align}
	P &= \frac{Ak_{W0}}{k_{R0}+k_{W0}} \frac{4}{3}\pi \int_{r_0}^{r_c} \mathrm{d}r~ r^3 \left(1-e^{-\frac{k_{R0}+k_{W0}}{r^6}t}\right)  \nonumber \\
	&= \frac{Ak_{W0}}{k_{R0}+k_{W0}} \frac{4}{3}\pi
    \frac{1}{4} \left[r^4 - \frac{2}{3}\left(\right(k_{R0}+k_{W0}\left) t\right)^{2/3} \Gamma\left(-\frac{2}{3},\frac{k_{R0}+k_{W0}}{r_c^6}t\right)\right]_{r_0}^{r_b} +C \label{eq:IncompleteGamma_Solution}
\end{align} 
with $\Gamma$ being the incomplete Gamma function.
$r_0$ and $r_c$ are a lower and upper cut-off radius around a paramagnetic center.
The lower bound describes the spins that are invisible to NMR owing to their large hyperfine shift (quenched spins) and $r_C$ would describe the Wigner-Seitz radius.
$C$ is the constant of integration and chosen as zero in the following.
Since $k_{W0}$, $k_{R0}$, $r_0$, $r_c$ and $t$ being positive real numbers, this solution is valid. 
Since our polarization is positive and our time is a positive real number, we pick the real part of the lower incomplete Gamma function as our solution for the polarization build-up. 
We choose $r_0=0$ as the number of quenched spins is assumed to small compared to the total number of spins ($r_0<<r_C$).
Furthermore, we choose $r_c=1$ as this ensures that $P\in [0,1]$.
This can be interpreted as $r_c$ defining the characteristic length scale of the system.
The resulting build-up curve is fitted with a stretched exponential of the form
\begin{align}
	P = P_{0,s} \left(1-e^{-\left(t/\tau_s\right)^\kappa}\right) \label{eq:StretchExp_SI}
\end{align}  
with $P_{0,s}$ being the steady-state polarization of the stretched exponential build-up, $\tau_s$ the build-up time constant and $\kappa$ the stretch exponent. 
Simulated build-ups of equation \eqref{eq:IncompleteGamma_Solution} for $r_c=1$ can be accurately fitted with a stretched exponential (Eq.~\eqref{eq:StretchExp_SI}) for a large range of $k_{W0}+k_{R0}$ with an example for this given in Fig.~\ref{fig:Simulation_stretchExp}a. 
In Fig.~\ref{fig:Simulation_stretchExp}b and c, the dependence of the stretched exponential build-up parameters based on $k_{W0}+k_{R0}$ is shown.
The stretched exponential build-up time $\tau_\mathrm{s} \approx 3.229 \left(k_{W0}+k_{R0} \right)^{-1}$ as extracted from the fit in Fig.~\ref{fig:Simulation_stretchExp}b.

The corresponding polarization decay/ relaxation after stopping the MW irradiation is straightforwardly derived from Eq.~\eqref{eq:IncompleteGamma_Solution} through replacing the $\left(1-e^{-\frac{k_{R0}+k_{W0}}{r^6}t}\right)$ with $e^{-\frac{k_{R0}}{r^6}t}$ in the integral. 
This cancels the $r_c^4$ term in the solution. 
The corresponding experimental model would be $P = P_{0,s} e^{-\left(t/\tau_s\right)^\kappa}$.

All the experimentally measured exponents reported in Ref.~\cite{jardon-alvarez_enabling_2020} exponents are between 0.69 and 0.74.
These values are slightly larger than the exponents shown in Fig.~\ref{fig:Simulation_stretchExp}c, although they are within the uncertainty limits of the experimentally measured exponents. 
If a systematic difference between the measured and simulated exponents exists, two possible explanations could be the presence of weak spin diffusion in the experiment or the assumption of an angle-independent hyperfine coupling. 
It would be interesting to investigate if a weak spin diffusion would lead to a larger exponent, eventually reaching $\kappa=1$ if spin diffusion in the system is significant.
If this would be the case, the discrepancy between the simulated (cf. Fig.~\ref{fig:Simulation_stretchExp}c) and an eventually measured exponent would indicate the balance between direct and spin diffusion mediated hyperpolarization transfer.

\clearpage

\end{document}